\documentclass[lettersize,journal]{IEEEtran}
\usepackage{amsmath,amssymb,amsfonts}
\usepackage{algpseudocode}
\usepackage{array}
\usepackage[caption=false,font=normalsize,labelfont=sf,textfont=sf]{subfig}
\usepackage{textcomp}
\usepackage{stfloats}
\usepackage{url}
\usepackage{verbatim}
\usepackage{graphicx}
\usepackage{acronym}
\usepackage[capitalize]{cleveref}   
\usepackage{acronym}
\usepackage{multirow}
\usepackage{siunitx}
\usepackage{booktabs}
\usepackage{tikz}
\usetikzlibrary{backgrounds, matrix, shapes.geometric, shapes.arrows, positioning, calc, arrows, arrows.meta, decorations.markings, decorations.pathreplacing, shapes, fit, automata, patterns, 3d, patterns.meta}
\usepackage{graphicx}
\usepackage{enumitem}
\usepackage{algorithm}
\usepackage{pgfplots}
\pgfplotsset{compat=1.18}
\usepgfplotslibrary{groupplots, fillbetween}
\usepackage{pgfplotstable}  
\pgfplotsset{table/search path={.,Figures}} 
\usepackage{standalone}
\usepackage{cite}
\usepackage{mycolors}
\usepackage{pgfplots}
\usepackage{xcolor} 
\usepackage{booktabs} 
\usepackage{multirow} 
\usepackage{makecell}

\def\BibTeX{{\rm B\kern-.05em{\sc i\kern-.025em b}\kern-.08em
    T\kern-.1667em\lower.7ex\hbox{E}\kern-.125emX}}
\usepackage{balance}

\tikzset{
  portgroup/.pic={
    \node[single arrow, draw=darkgray, fill=white,
      minimum height=0.5cm,
      single arrow head extend=2pt,
      yscale=0.5
    ] (write) at (0,0) {};

    \node[single arrow, draw=darkgray, fill=white,
      minimum height=0.5cm,
      single arrow head extend=2pt,
      yscale=0.5,
      rotate=180
    ] (read1) at ([xshift=0.12cm, yshift=0.2cm]write.north) {};

    \node[single arrow, draw=darkgray, fill=white,
      minimum height=0.5cm,
      single arrow head extend=2pt,
      yscale=0.5,
      rotate=180
    ] (read2) at ([yshift=0.35cm]read1.north) {};
  }
}

\newif\ifSHOWREVISIONS
\SHOWREVISIONSfalse

\definecolor{revR1}{RGB}{176,140,0}
\definecolor{revR2}{RGB}{0,140,60}
\definecolor{revR3}{RGB}{220,110,0}

\ifSHOWREVISIONS
  \newcommand{\highlightRI}[1]{{\color{revR1}#1}}
  \newcommand{\highlightRII}[1]{{\color{revR2}#1}}
  \newcommand{\highlightRIII}[1]{{\color{revR3}#1}}
\else
  \newcommand{\highlightRI}[1]{#1}
  \newcommand{\highlightRII}[1]{#1}
  \newcommand{\highlightRIII}[1]{#1}
\fi

\ifSHOWREVISIONS

\else

\fi

\acrodef{AI}{artificial intelligence}
\acrodef{API}{application programming interface}
\acrodef{CNN}{convolutional neural network}
\acrodef{CP}{constrained programming}
\acrodef{DAE}{decoupled access-execute}
\acrodef{DDR}{double data rate}
\acrodef{DM}{data movements}
\acrodef{DMA}{direct memory access}
\acrodef{DRAM}{dynamic random access memory}
\acrodef{DSE}{design space exploration}
\acrodef{EDP}{energy-delay product}
\acrodef{eTOPS}{effective teraoperations per second}
\acrodef{FPGA}{field-programmable gate array}
\acrodef{GPU}{graphical processing unit}
\acrodef{HDA}{heterogeneous dataflow accelerator}
\acrodef{HW}{hardware}
\acrodef{HW/SW}{hardware-software}
\acrodef{ILP}{integer linear programming}
\acrodef{IR}{intermediate representation}
\acrodef{LLM}{large language model}
\acrodef{LTP}{latency-TOPS product}
\acrodef{MAC}{multiply-and-accumulate}
\acrodef{MIP}{mixed-integer programming}
\acrodef{ML}{machine learning}
\acrodef{NN}{neural network}
\acrodef{NoC}{network-on-chip}
\acrodef{NPU}{neural processing unit}
\acrodef{eNPU}{embedded NPU}
\acrodef{iNPU}{integrated NPU}
\acrodef{MPU}{micro-processor-unit SoC}
\acrodef{OCM}{on-chip memory}
\acrodef{PTQ}{post traininq quantization}
\acrodef{SoC}{system-on-chip}
\acrodef{SOTA}{state-of-the-art}
\acrodef{SRAM}{static random access memory}
\acrodef{SW}{software}
\acrodef{TCM}{tightly-coupled memory}
\acrodef{TOPS}{teraoperations per second}
\acrodef{V2P}{virtual-to-physical}

\begin{document}

\title{eIQ Neutron: Redefining Edge-AI Inference with Integrated NPU and Compiler Innovations}

\author{
    Lennart Bamberg\IEEEauthorrefmark{1},
    Filippo Minnella\IEEEauthorrefmark{1},
    Roberto Bosio\IEEEauthorrefmark{1}~\IEEEmembership{Student Member,~IEEE},
    Fabrizio Ottati,
    Yuebin Wang,
    Jongmin Lee,
    Luciano Lavagno~\IEEEmembership{Senior Member, IEEE},
    Adam Fuks
    \thanks{\IEEEauthorrefmark{1} Authors contributed equally to this work.}
    \thanks{Lennart Bamberg,  Yuebin Wang, Jongmin Lee, and Adam Fuks are with NXP Semiconductors. 
    Filippo Minnella is with Nvidia.
    Roberto Bosio and Luciano Lavagno are with Politecnico di Torino, Italy.}
}

\markboth{eIQ Neutron: Redefining Edge-AI Inference with Integrated NPU and Compiler Innovations}{}

\maketitle

\begin{abstract} 
Neural Processing Units (NPUs) are key to enabling efficient AI inference in resource-constrained edge environments. 
While peak tera operations per second (TOPS) is often used to gauge performance, it poorly reflects real-world performance
and instead typically correlates with higher silicon cost. 
To address this, architects must focus on maximizing compute utilization, without sacrificing flexibility. This paper presents the eIQ Neutron efficient-NPU---integrated into a commercial flagship MPU—alongside co-designed compiler algorithms. The architecture employs a flexible, data-driven design, while the compiler uses a constrained programming approach to optimize compute and data movement based on workload characteristics.
Compared to the leading embedded NPU and compiler stack, our solution achieves an average speedup of 1.8$\times$ (4$\times$ peak) at equal TOPS and memory resources across standard AI benchmarks. Even against NPUs with double the compute and memory resources, Neutron delivers up to 3.3$\times$  higher performance. 

\end{abstract}


\section{Introduction}
\label{sec:intro}

Cloud-based \ac{ML} inference incurs significant latency and power overhead due to data transfers between user devices and remote servers. It also raises privacy and availability concerns, particularly in mission-, security-, or safety-critical applications. Consequently, there is growing interest in performing efficient \ac{ML} inference directly on edge devices, both in industry~\cite{executorch,david2021tensorflow} and academia~\cite{lin2020mcunet,defines}.

Edge devices are typically resource-constrained, necessitating domain-specific accelerators such as \acfp{NPU}, supported by dedicated compiler frameworks~\cite{dally2023hardware,Dally2020DomainSpecific,arm_ethos_u85,coral_dev_board,hailo_15_ai_vision_processor}. These \acp{NPU} enable compute-intensive workloads within tight power and memory budgets.

\highlightRI{
\Ac{HW/SW} co-design for \acp{NPU} has proven effective in reducing model complexity and improving the deployability of modern \ac{ML} models on resource-constrained systems~\cite{deng2020model,jacob2018quantization,kuzmin2022fp8}. Nevertheless, state-of-the-art edge-AI models remain computationally demanding in terms of \ac{MAC} operations. Consequently, \acp{NPU} are commonly characterized by their peak \ac{TOPS}\footnote{Throughout this work, we follow the convention that one \ac{MAC} counts as two operations---one multiplication and one addition---such that peak \ac{TOPS} is defined as twice the number of \ac{MAC} units multiplied by the clock frequency and divided by $10^{12}$.}. However, peak \ac{TOPS} is a poor proxy for real-world performance because the provisioned \ac{MAC} units may remain substantially underutilized during inference. Moreover, by definition, increasing peak \ac{TOPS} requires either more \ac{MAC} units---together with the associated accumulator, register, and interconnect resources---or a higher operating frequency. Under otherwise comparable implementation conditions, both approaches carry corresponding area and/or power costs~\cite{reuther2022ai,horowitz2014computing,sze2020evaluate}. This consideration is particularly relevant for accuracy-critical applications in which aggressive approximations, such as analog \ac{MAC} blocks or ultra-low-bitwidth quantization, may be unacceptable.

\cref{tab:npus_eTOPS} illustrates the resulting gap between provisioned peak compute and delivered application performance for two industry-leading edge \acp{NPU}: an up-to-\SI{4}{TOPS} \ac{eNPU}~\cite{arm_ethos_u85}, which can be integrated into different SoC platforms, and an up-to-\SI{11}{TOPS} \ac{iNPU}, in a dedicated AI Vision Processor~\cite{hailo_15_ai_vision_processor}. Effective \ac{TOPS}, defined as the ratio of executed operations to inference latency, shows that the delivered throughput of both \acp{NPU} is substantially below their respective advertised peak throughput (by up to $42\times$). Notably, despite its substantially higher peak \ac{TOPS}, the \SI{11}{TOPS} \ac{iNPU} does not outperform the \SI{4}{TOPS} \ac{eNPU} on these benchmarks.

\begin{table}[t]
     \footnotesize
     \centering
     \caption{Effective \ac{TOPS} of commercial edge \acp{NPU}.}
     \label{tab:npus_eTOPS}
     \begin{tabular}{lccc}
         \toprule
         \textbf{NPU} & \textbf{Peak \ac{TOPS}} & \textbf{ResNet50 V1} & \textbf{EfficientNet Lite0} \\
         \midrule
         \acs{eNPU} & 4 & 0.73 & 0.82 \\
         \acs{iNPU} & 11 & 0.89 & 0.26 \\
         \bottomrule
     \end{tabular}
     \vspace{-4.2mm}
\end{table}

Therefore, peak \ac{TOPS} alone should be complemented by metrics that capture actual compute utilization under realistic memory and bandwidth constraints and reward designs that achieve a target latency with less provisioned peak compute. This distinction is particularly important because data movement between compute blocks and across the memory hierarchy, both on- and off-chip, incurs substantial performance, area, and power costs~\cite{sze2020evaluate,dally2023hardware}.
}

In-memory compute architectures reduce these penalties, but require sufficient on-chip memory to store entire models—impractical for modern, large-scale \ac{ML} models or multi-model deployments  in mission-, safety-, or security-critical systems where accuracy and redundancy are essential. Consequently, commercial products favor near-memory-compute \acp{NPU} with small, tightly coupled memories and integrated data-mover engines. These architectures enable efficient execution of diverse and complex edge-AI workloads that reside in off-chip DDR or flash memory, offering greater flexibility and scalability than in-memory compute solutions~\cite{arm_ethos_u85,hailo_15_ai_vision_processor}.

In this work, we introduce a novel near-memory-compute \ac{NPU} architecture, called \textit{eIQ Neutron}, and a scalable compilation strategy that jointly optimize hardware and software for efficient edge inference, maximizing performance per peak-TOPS and thereby cost-efficiency. While the \ac{NPU} and software stack are designed for broad applicability—from low-cost MCUs to high-end MPUs—this paper focuses on a \SI{2}{TOPS} implementation integrated into a commercial flagship MPU\@. The resulting hardware–software system achieves superior efficiency--defined as inference latency relative to available compute---across standard computer-vision benchmarks compared to reference platforms. 

The remainder of this paper is organized as follows: \cref{sec:rel_work} summarizes related work; \cref{sec:npu_arch} describes the \ac{NPU} architecture; \cref{sec:sw_stack} details the software stack and compiler; \cref{sec:results} presents experimental results; and \cref{sec:conclusions} concludes the paper.

\section{Related Work}
\label{sec:rel_work}

Efficient edge inference is constrained less by peak MAC throughput than by data movement under tight on-chip memory budgets. Moving tensors---especially to and from off-chip memory---can cost orders of magnitude more energy and cycles than arithmetic, making performance gains dependent on minimizing traffic and maximizing on-chip reuse~\cite{sze2017efficient}. This observation has shaped most modern accelerator architectures and compiler strategies for the edge.

Within these constraints, \ac{HW/SW} co-design is essential. Efficient inference requires joint decisions across the stack: from neural network architecture and quantization schemes to compiler-level optimizations such as operator fusion, tiling, scheduling, and strategic tensor placement within limited SRAM. These transformations expose and exploit the underlying hardware’s dataflow model and memory hierarchy.

\highlightRII{
Architectural research has pursued efficient edge inference through reconfigurable dataflows, sparsity-aware execution, and precision scaling~\cite{eyeriss,zena,Sim,Park}. In particular, dot-product-based acceleration has long been employed for embedded convolutional workloads. Desoli \textit{et al.}~\cite{desoli2017dcnn} and Conti \textit{et al.}~\cite{conti2017iot} integrated convolution acceleration with programmable processing and local memories in embedded SoCs. More recently, Prasad \textit{et al.}~\cite{prasad2023specialization} combined a configurable DNN accelerator with programmable cores in a heterogeneous cluster targeting AR/VR workloads. Islamoglu \textit{et al.}~\cite{islamoglu2023ita} further demonstrated how dataflow specialization can reduce intermediate traffic by fusing quantized attention and softmax.
}

\highlightRII{
In parallel, compiler research has addressed how neural-network graphs are transformed into execution plans that exploit an accelerator's dataflow while satisfying its memory and hardware constraints. Heuristic and design-space-exploration frameworks such as DNNVM, DeFiNES, and TANGRAM optimize graph transformations and hardware mappings~\cite{DNNVM,defines,tangram}. ZigZag~\cite{mei2021zigzag} and QuickFlow~\cite{ronzani2025quickflow} explore layer-level loop ordering, tiling, and processing-element allocation for spatial accelerators. Deeploy~\cite{scherer2024deeploy} and MATCH~\cite{hamdi2025match} address memory-aware deployment on heterogeneous edge platforms. Formal approaches, including CoSA, Klotski, and Stream, formulate mapping, scheduling, placement, and layer fusion as optimization problems~\cite{cosa,klotski,stream-dse}.
}

\highlightRII{
The works discussed above span a broad range of architectural and compiler techniques, from individual mechanisms to integrated accelerator designs and deployment frameworks. Complementing this research perspective, commercial \acp{NPU} provide insight into how multiple architectural mechanisms are combined under practical system constraints. We therefore next discuss representative commercial designs relevant to the architectural context of this work.
}

\textbf{Micro-NPUs for deeply embedded systems:} These accelerators integrate tightly with application processors and microcontrollers and rely on compiler optimizations such as parameter compression and layer fusion to reduce SRAM usage, while recent iterations add structured sparsity and mixed-precision support~\cite{arm_ethos_u85}.

\textbf{High-throughput systolic arrays:} These designs employ INT8 systolic arrays with on-chip SRAM as a software-managed cache for model parameters. Performance degrades when models exceed this budget, requiring segmentation or pipelined execution across multiple devices~\cite{coral_dev_board}.

\textbf{Dataflow-centric fabrics:} These architectures adopt a distributed-memory fabric. The corresponding compiler partitions DNN graphs across this fabric, explicitly scheduling computation and buffer placement to exploit locality and sustain utilization~\cite{hailo_15_ai_vision_processor}.

\textbf{Reconfigurable NPUs with sparsity support:} Recent proposals integrate sparsity-aware, reconfigurable NPUs within a flagship mobile SoC, combining fine-grained structured sparsity with flexible dataflows for high utilization and energy efficiency~\cite{samsung_npu}. 

\highlightRIII{
Across these designs, efficient edge inference relies on tight
hardware--software coupling to coordinate computation, data movement,
and memory use at fine granularity. Within this established design
space, Neutron contributes a near-memory architecture-compiler
co-design in which the hardware organization and compiler optimizations
are jointly designed to sustain compute utilization under constrained
local-memory, interconnect, and off-chip-bandwidth resources. At the
architectural level, parallel dot-product engines are organized around
shared operand streams, multiple local accumulators, programmable
multidimensional data access, and banked software-managed local memory.
At the compiler level, architecture-specific tensor formats and spatial
mappings are combined with constraint-programming formulations for
temporal tiling and layer fusion, latency-hiding scheduling of compute
and data-transfer jobs, and bank-aware memory allocation. The evaluated
four-core, 2-TOPS subsystem constitutes a commercial instantiation of
this co-design and is validated end to end on production silicon across
image-classification, object-detection, and semantic-segmentation
workloads.
}

\section{NPU architecture}
\label{sec:npu_arch}

In this section, we will outline the architecture of our \textit{eIQ Neutron} \ac{NPU}\@. We start with the principles by which the system is designed before we outline the architectural details.

\subsection{Architectural Principles}
A sound processor architecture should be guided by first principles. Below, we outline the key principles derived for designing resource-efficient, domain-specific accelerators in aggressively scaled technology nodes (16\,nm and below).
Adhering to these technology-aware principles enables higher performance at lower cost.

\textbf{1) Interconnect-Centric Design:} In advanced CMOS nodes, gate capacitances shrink with transistor size, whereas metal-wire capacitances increase with scaling as parasitic coupling capacitances  grow with wire density. For example, transferring 256\,bits over a 1\,mm interconnect in 10\,nm consumes more energy than hundreds of integer \ac{MAC} operations~\cite{dally2024throughput}. Interconnects dominate placement density, timing closure, and thus all PPA metrics. Architectures must therefore minimize the global and intermediate interconnect bandwidth requirements for full \ac{MAC} utilization and avoid arbitration loops.


\textbf{2) Memory Locality and Latency Tolerance:} Memory scaling outpaces interconnect scaling but lags transistor scaling. Accessing local SRAM with high bandwidth is acceptable (register-level locality is ideal), whereas higher-level memory—especially off-chip DRAM—must be minimized, as resource-efficient systems cannot afford high DDR bandwidth. Architectures should: tolerate latency via deep pipelining on long interconnect paths to achieve high throughput; and minimize global data movement by high data locality achieved at low local memory capacities.

\textbf{3) Instruction Overhead Reduction:} Programmability remains essential for future-proofing in the rapidly evolving \ac{AI} domain. However, modern edge inference relies on low-cost, low-bitwidth (8–16\,bit) integer arithmetic, where a single RISC instruction fetch, decoding, operand fetch, and result write back has magnitudes higher costs 
than the core \ac{MAC} operation~\cite{horowitz20141}.  Minimizing these instruction and register accesses overhead per operation demands data-driven, systolic architectures, capable of performing large arithmetic sets per programming step.

\subsection{Neutron Core Architecture}
Our accelerator employs a dot-product systolic architecture. Traditional high-throughput systolic arrays rely on numerous wide (32-bit) accumulators---introducing significant wire, logic, and register overhead---and suffer from poor utilization on small workloads. Our dot-product structure eliminates these limitations.

\begin{figure}
    \centering
    \includegraphics[width=0.8\columnwidth]{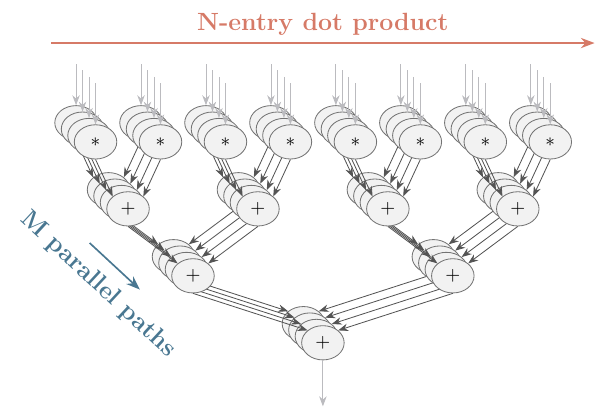}
    \caption{Core dot-product array.}
    \label{fig:dpe}
\end{figure}

The core consists of $M$ parallel, pipelined, dot-product units, each completing a dot-product of two vectors of length $N$ per cycle (\textit{cf}. \cref{fig:dpe}). Thus, it enables $2NM$ operations per cycle. To reduce bandwidth needs, all $M$ units share one operand vector, lowering the required input bandwidth for full utilization from $NM$ (for int8) to $N$ bytes, provided sufficient data parallelism exists. An exception occurs when one operand can remain fully stationary, allowing high utilization even with distinct weights across the parallel units after a short initialization phase. 

The compute flow is output-stationary to completely avoid outside memory accesses for wide 32-bit accumulator values. Additionally, $A$ parallel accumulator values per dot-product unit---stored in small local scratchpad memories---enable computing multiple dot-product results concurrently while reusing the second operand. This reduces the bandwidth needs also for the second, non-shared operand by up to a factor $A\times$.
Pipelines support 8-bit by 16-bit dot-products using 8-bit multipliers with a two-cycle decomposition, requiring adder trees of 24-bit input and 27-bit output, instead of 16-bit and 19-bit respectively.

Further bandwidth reduction is achieved through a \emph{data engine}: a programmable, word-level pre-fetcher with a multi-dimensional address generation loop fetching in a  2D register file, with support for byte-level scrolling, feeding the dot-product engines. It enables spatial and temporal data reuse, exploiting spatial and temporal locality in \ac{AI} layers. An optional local scratchpad of size $W_C$ leverages shift invariance in embedding and convolution operations: parameters for the first $M$ pixels/tokens are fetched from external memory, then reused from the cache. If parameters exceed $W_C$, the layer is partitioned into smaller sub-problems with fewer output features, or the remaining parameters are streamed per set.

Final 32-bit accumulation results can be rescaled to 8 or 16 bits and passed through a dedicated activation engine that applies arbitrary nonlinear functions (e.g., ReLU, Swish, Mish) and supports on-the-fly min/max pooling before writing outputs back to memory. This fusion of operations into a single pipeline further reduces memory bandwidth requirements.

Overall, interconnect and bandwidth demands can be reduced at all ends by tuning $M$, $A$, or $W_C$. Increasing $M$ incurs no local memory, just logic, cost, while $A$ and $W_C$ add minimal scratchpad overhead. Deep pipelining and configurable outstanding transactions on operand and result buses enable high performance even under high interconnect latency.
To minimize programming overhead, the memory-mapped accelerator can execute millions of operations in a single programming step. Next-task programming can overlap with execution to hide latency.

In this work, we configure $N=M=16$, achieving 0.5\,TOPS at 1\,GHz per engine with only three 128-bit buses. To further reduce programming overhead and bus utilization at minimal hardware cost, we set $A=2M$ and $W_C=8\,\mathrm{kB}$.

\subsection{Neutron Multi-Core System}

\begin{figure}
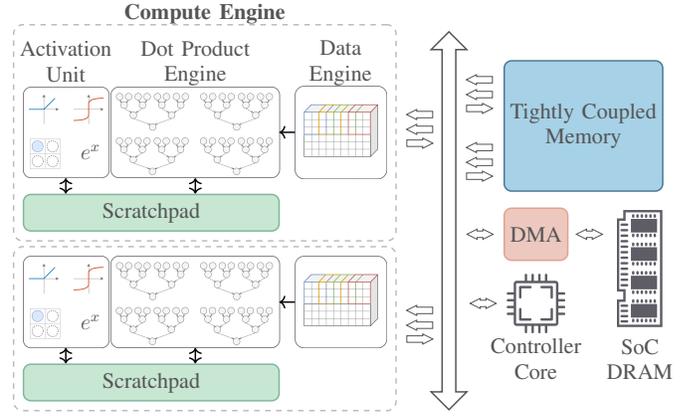

    \centering
    \includestandalone[width=1.0\linewidth]{Figures/npu}
    \caption{Proposed NPU subsystem integrated into SoCs.}
    \label{fig:npu_subsystem}
\end{figure}

\cref{fig:npu_subsystem} illustrates the architecture of the entire \ac{NPU} subsystem, integrated into several commercial SoCs as an AXI–interconnected accelerator IP\@. The subsystem comprises a configurable number of the above-described compute cores (four in the \ac{NPU} considered in this work, delivering 2\,TOPS), connected via a multilayer bus to a \ac{TCM}, a system data mover, and a RISC-V controller core.

\textbf{Tightly Coupled Memory (TCM):} The \ac{TCM} is a software-managed on-chip memory located near the compute engines, providing high-bandwidth access for neural network data and parameters. It is organized into multiple banks to enable high bandwidth.
\highlightRI{
To close timing in physical design at high performance and low implementation cost, the banks are non-arbitrated, replacing costly arbiters and their one-hot acknowledgment signals with OR gates without return-signal routes. Consequently, bank-access conflicts must be avoided by the compiler. A \ac{V2P} translation table enables dynamic address-to-bank remapping in idle mode.
The number and size of the banks are architectural parameters that trade off compiler flexibility against implementation cost. Fewer banks tighten the compiler constraints but reduce memory-peripheral overheads. For the targeted commercial 16-nm technology, 16 16-kB banks per compute engine provide sufficient flexibility while keeping peripheral overheads small.
}

\textbf{Controller and Data Movement:} The subsystem includes a lightweight RISC-V controller and a \ac{DMA} engine for data exchange between the \ac{TCM} and the rest of the system (e.g., DDR). The \ac{DMA} supports multi-dimensional strided transfers and TCM-to-TCM data rearrangements, enabling efficient tiling and format transformations (see \cref{subsec:format_and_tiling}).

\highlightRIII{
The architecture uses a single data mover with one channel. Multiple DMA channels would increase interconnect, arbitration, and implementation costs, while the targeted AI workloads permit large, statically scheduled sequential DDR transfers that efficiently utilize the available bandwidth. Consequently, multiple data-movement jobs are serialized and scheduled back-to-back by the controller core.
}

\textbf{Bandwidth and Control Optimization:} To minimize control overhead, avoid Amdahl’s law bottlenecks, and reduce memory bandwidth and capacity requirements a multi-layer bus system  supports operand sharing across compute engines. The controller can configure operand buses in a sharing mode, where one operand stream (data or parameters) is broadcast from the \ac{TCM} to all engines. This requires engines to operate in lockstep on equally sized layer partitions (with optional padding). Programming can be applied globally or per engine, reducing control overhead while maintaining~flexibility.
\section{Software stack}
\label{sec:sw_stack}

The \ac{NPU} relies on a library of optimized embedded-C kernels implementing layers commonly found in modern \ac{ML} models, similar in concept to CMSIS-NN or the ATen library~\cite{ATen}. The compiler generates an executable composed of compute jobs, data-transfer jobs, and synchronization barriers for the internal RISC-V controller. Compute jobs invoke the kernel library to program the \ac{AI} compute cores, while data transfers and synchronization rely on firmware that configures the \ac{NPU}'s \ac{DMA} engine and monitors execution status.

The compilation flow follows a standard structure: a frontend ingests the neural network model and converts it into an internal \ac{IR}; the mid-end lowers and optimizes the graph through tiling, scheduling, and memory allocation; and the backend emits code based on the operator library and firmware, which is compiled into an executable. The host runtime and frontend leverage LiteRT~\cite{litert} (formerly TensorFlow Lite) for model parsing and fallback execution of unsupported operators, though the mid-end, backend, library, and firmware remain framework-agnostic.

\highlightRIII{In the remainder of this section, we focus on the novel contribution: how the mid-end leverages \ac{CP} based optimization to schedule and allocate resources for the model graph, maximizing utilization under the constraints of the proposed \ac{NPU} architecture.}


\subsection{Format Selection and Tiling} \label{subsec:format_and_tiling}


\Acl{ML} models are a good fit for highly-parallel architectures due to the
inherent parallelism in the compute tasks. A convolution layer in modern
\acp{NN} takes two input tensors: an activation tensor, named $ifmap$, of shape
($inpH,$ $inpW$, $inpC$); and a weight and bias tensor, named $parameters$, of
shape ($outC$, $filterH$, $filterW$, $inpC$).  It produces an output tensor,
named ${ofmap}$, of shape ($outH$, $outW$, $outC$). The computation procedure
for generating this output tensor is presented in \cref{alg:conv}.
\begin{algorithm}[t]
\footnotesize
\caption{Convolution. Nested loops are inlined.}
\label{alg:conv}
\begin{algorithmic}[1]
    \For{$h_o \in [0,~\mathit{outH}),~w_o \in [0,~\mathit{outW}),~c_o \in [0,~\mathit{outC})$}
        \State $a \leftarrow 0$
        \For{$h_f \in [0,~\mathit{filterH}),~w_f \in [0,~\mathit{filterW}),~c_i \in [0,~\mathit{inpC})$}
            \State $h_i \leftarrow h_o*S+h_f$ \Comment{$S$: stride.}
            \State $w_i \leftarrow w_o*S+w_f$
            \State $a \leftarrow a + parameters[c_o,~h_f,~w_f,~c_i]~*~ifmap[c_i,~h_i,~w_i]$
        \EndFor
        \State $ofmap[h_o,~w_o,~c_o] \leftarrow a$
    \EndFor
\end{algorithmic}
\end{algorithm}

\textbf{Spatial Tiling:} Spatial tiling describes how the computation of a given \ac{NN} layer is distributed over multiple compute engines. Algorithm \ref{alg:conv} reveals that each of the three output dimensions provides parallelism exploitable by spatial tiling.

When spatially tiling along the channel dimension, $outC$, the same input activations are shared across different filters---increasing efficiency---with each compute engine processing a unique subset of the parameters. We refer to this as \emph{depth parallelism}. Assuming an architecture with $N$ compute engines, as described in Section~\ref{sec:npu_arch}, \emph{depth parallelism} is illustrated in Algorithm~\ref{alg:depth_par} using NumPy-style~\cite{harris2020array} tensor slicing syntax.

\begin{algorithm}[t]
\footnotesize
\caption{Data distribution among engines in depth parallelism.}
\label{alg:depth_par}
\begin{algorithmic}[1]
    \renewcommand{\algorithmicfor}{\textbf{parfor}}
    \For{$n \in [0,~N)$}
        \State $c_{o_{start}} \leftarrow n * \frac{outC}{N}$
        \State $c_{o_{end}} \leftarrow (n + 1) * \frac{outC}{N}$
        \State $engines[n] \leftarrow ifmap[\colon,~\colon,~\colon]$ 
        \State $engines[n] \leftarrow parameters[c_{o_{start}}\colon c_{o_{end}},~\colon,~\colon,~\colon]$ 
        \State $engines[n] \rightarrow ofmap[\colon,~\colon,~c_{o_{start}}\colon c_{o_{end}}]$
    \EndFor
\end{algorithmic}
\end{algorithm}
A key constraint for lockstep execution across compute engines is that each engine must write to a separate memory bank to prevent access conflicts. Hence, $ofmap$ is fragmented in memory along $outC$. This means that $ifmap$, generated by a previous compute job as $ofmap$, is fragmented along $inpC$. Consequently, when broadcasting $ifmap$ to the engines and processing it depth-first, the addressing sequence rotates among the fragments in memory at the word level, as the considered tensor format for compute is \textit{HWC}\@.
This approach eliminates the need for an explicit data rearrangement step to make $ofmap$ contiguous in memory for further processing.

The alternative to depth parallelism is to tile by one of the space coordinates, $outH$ or $outW$. 
Due to the equivalence of the two variants, we limit our spatial tiling to only consider tiling by $outH$ and refer to it as \emph{line parallelism}.
With line-parallel computing, different compute engines work on different output lines but the same output channels. Since parameters and biases in a convolutional operation are independent of the spatial coordinates (enabling translation invariance), \emph{line parallelism} allows the parameters to be shared across the engines to increase efficiency.
The locality of the convolution operation requires each compute engine to access different input data windows, which depend on factors like the convolution stride and kernel height.
The data distribution for \emph{line parallelism} is shown in \cref{alg:line_par}.
\begin{algorithm}[t]
\caption{Data distribution among engines in line parallelism.}
\footnotesize
\label{alg:line_par}
\begin{algorithmic}[1]
    \renewcommand{\algorithmicfor}{\textbf{parfor}}
    \For{$n \in [0,~N)$}
        \State $h_{o_{start}} \leftarrow n * \frac{outH}{N}$
        \State $h_{o_{end}} \leftarrow (n + 1) * \frac{outH}{N}$
        \State $h_{i_{start}} \leftarrow h_{o_{start}} * S$
        \State $h_{i_{end}} \leftarrow h_{o_{end}} * S  + filterH -1$
        \State $engines[n] \leftarrow ifmap[h_{i_{start}} \colon{}h_{i_{end}},~\colon,~\colon]$ 
        \State $engines[n] \leftarrow parameters[\colon,~\colon,~\colon,~\colon]$ 
        \State $engines[n] \rightarrow ofmap[h_{o_{start}}\colon h_{o_{end}},~\colon,\colon]$
    \EndFor
\end{algorithmic}
\end{algorithm}

To avoid bank access conflicts in line-parallel compute, while achieving lockstep compute, the input windows for each engine must be located in mutually exclusive memory banks.
However, when the convolution kernel height exceeds one, these input windows overlap across engines, as illustrated in ~\cref{alg:line_par}.
The previous compute, which produced the required $ifmap$ as $ofmap$, already wrote the feature map to different banks to avoid access conflicts on result writing.
Therefore, before computation can begin, overlapping input regions must be copied between memory banks using \ac{TCM}-to-\ac{TCM} transfers. The amount of data to be copied depends on the input-tensor dimensions and convolution properties. 

Depthwise convolutions under \emph{depth parallelism} form a special case in which every engine only needs the input channels for the respective output channels it works on. Thus, all engines can work with one of the mutually exclusive subsets of the inputs produced in the previous step without any multi-fragment collection or pre-compute copying. 

Fully connected layers and matrix-matrix multiplications are handled as 1$\times$1 convolutions, while element-wise operations like addition and Hadamard layers are interpreted as paired depthwise computations. Scalar operations, such as additions with constants, are similarly treated as 1$\times$1 depthwise operations. Modern Transformer-based neural networks are implemented by considering the embedding dimension as $C$ and the token dimension as $H$ to obtain the same efficiency gains from the two proposed tiling strategies. Thus, our tiling approach can be applied consistently across all modern \acp{NN}\@.

\begin{figure}
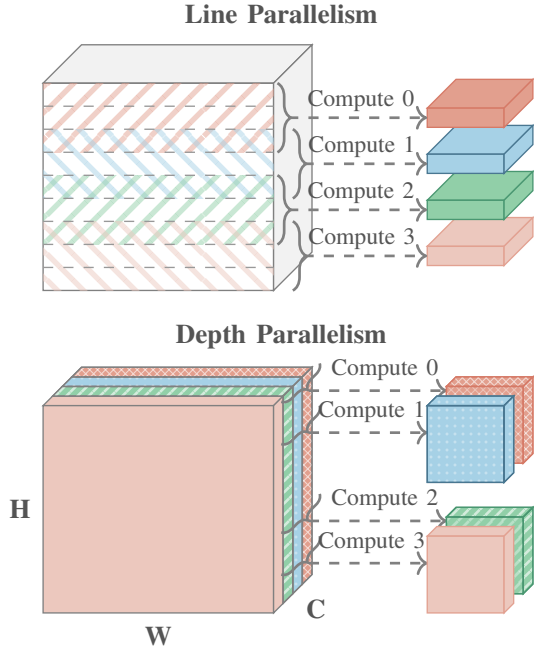

    \centering
    \includestandalone[width=0.8\linewidth]{Figures/formats}
    \caption{Data dependencies in depth and line parallelism.} 
    \label{fig:par_deps}
\end{figure}

\highlightRIII{
A visual summary of \emph{line} and \emph{depth parallelisms} is provided in
\cref{fig:par_deps}. When the number of tiled lines or channels is not evenly
divisible by the number of compute engines $N$, padding with garbage data is
applied to maintain lockstep execution. Furthermore, $ifmap$ and $ofmap$ fragments are
stored in \ac{TCM} with their channel dimension $C$ padded per memory fragment
to a multiple of the bus/word width, without increasing compute, to make all
transactions word aligned. For the evaluated NPU with four compute engines and
16-byte words, depth parallelism therefore permits padding-free execution when
the channel dimension is a multiple of 64, such that each of the four
fragments contains a multiple of 16 channels (i.e., multiple full words). For channel dimensions that are
only multiples of 16 but not 64---as found in some early \ac{CNN} layers---the compiler
can instead select \emph{line parallelism}, for which the channels are not
partitioned across engines and the lines can be evenly distributed across the
four engines. In the common neural networks, the output channel dimensions are always multiples of 16
and the corresponding line dimensions permit an even distribution across the
four engines when line parallelism is required. Hence, by selecting between these two
complementary formats, the compiler therefore avoids lockstep-compute- padding and memory-padding
overheads.

We refer to the memory layouts used by the compute jobs for \emph{line} and
\emph{depth parallelisms} as \emph{formats}.
Formats are crucial for optimizing compute utilization. The line-parallel
format is best suited for shallow layers with few channels but higher
resolution. Here, the depth-parallel format often cannot utilize all compute
engines. The contrary is true for layers with many channels. Not only because
\emph{depth parallelism} allows high compute utilization at low line-counts,
but also because it avoids pre-compute TCM copies. The compiler chooses the
most suitable format for each layer of the \ac{NN} by estimating execution
latencies and taking into account the overhead of switching formats between
consecutive layers, for which extra operators exist in the library.
}

\textbf{Temporal Tiling:} Once formats are assigned, the next step transforms the tensor-based graph into a tiled representation. Since feature maps may exceed the \ac{TCM} capacity, layers are divided into smaller tiles, each processed at different times. We refer to this process as temporal tiling. Determining an upper bound for tile size is straightforward: it is enough to ensure that the combined memory footprint of each tile and its required dependencies fits within the \ac{TCM}\@. However, such naive bounds are often suboptimal. A more detailed discussion of how temporal tiling is decided is provided in~\cref{subsec:opt}.

\subsection{Scheduling} \label{subsec:scheduling}
The scheduling problem involves converting a tile-based graph into a sequence of timed jobs, categorized into two types: data-transfer jobs (pushes, fetches, copies, \ac{V2P} update) and compute jobs (operator calls). While in literature scheduling often includes determining the tile computation order as well as the tile sizes, this work treats the two aspects separately. Here, scheduling assumes a given sequence of tiles and focuses solely on optimizing memory latency hiding, while ensuring compliance with all platform-specific constraints. A detailed discussion on the selection of an efficient computation order, which is referred to as layer fusion, can be found in \cref{subsec:opt}.

\textbf{Time Discretization:} The \ac{NPU} adopts a \ac{DAE} architecture, 
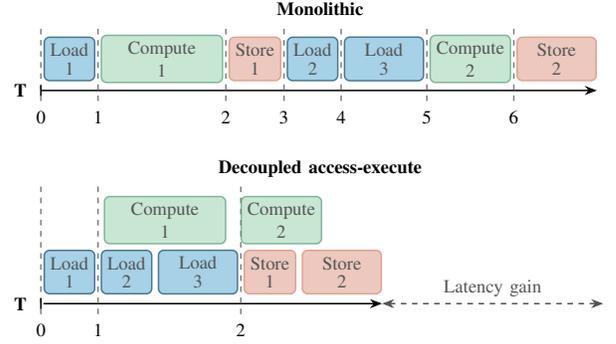
\begin{figure}
\centering
    \begin{tikzpicture}[
  stage/.style={rectangle, draw, minimum height=0.8cm, minimum width=1.5cm, align=center, rounded corners=1mm, thick},
  >=Stealth,
  node distance=0.2cm and 0.1cm
]

\node[stage, minimum width=0.5cm, fill=myblueRB, draw=myblueborderRB] (m1) {\textcolor{darkgray}{\shortstack{Load \\ 1}}};
\node[stage, minimum width=2.3cm, fill=mygreenRB!50, draw=mygreenborderRB!70, thick, right=of m1] (m2) {\textcolor{darkgray}{\shortstack{Compute \\ 1}}};
\node[stage, minimum width=0.5cm, fill=mypinkRB, draw=mypinkborderRB, thick, right=of m2] (m3) {\textcolor{darkgray}{\shortstack{Store \\ 1}}};

\node[stage, minimum width=0.5cm, fill=myblueRB, draw=myblueborderRB, thick, right=of m3] (m4) {\textcolor{darkgray}{\shortstack{Load \\ 2}}};
\node[stage, fill=myblueRB, draw=myblueborderRB, thick, right=of m4] (m4a) {\textcolor{darkgray}{\shortstack{Load \\ 3}}};
\node[stage, fill=mygreenRB!50, draw=mygreenborderRB!70, thick, right=of m4a] (m5) {\textcolor{darkgray}{\shortstack{Compute \\ 2}}};
\node[stage, fill=mypinkRB, draw=mypinkborderRB, thick, right=of m5] (m6) {\textcolor{darkgray}{\shortstack{Store \\ 2}}};

\node[align=center, yshift=0.5cm] (l1) at (current bounding box.north) {\textbf{Monolithic}};

\draw[->, thick] 
  ([yshift=-0.6cm, xshift=-0.05cm]m1.west) -- ([yshift=-0.6cm]m6.east);

\foreach \x in {m1, m2, m3, m4, m4a, m5, m6} {
  \draw[dashed, thick, gray] 
    ([yshift=-0.8cm, xshift=-0.05cm]\x.west) -- ([yshift=0.65cm, xshift=-0.05cm]\x.west);
}

\draw[thick] 
  ([yshift=-0.7cm, xshift=-0.05cm]m1.west) -- ++(0,0.2);

\node[left = of m1, xshift=-0.1cm, yshift=-0.6cm] {\textbf{T}};

\foreach \i/\x in {0/m1,1/m2,2/m3,3/m4,4/m4a,5/m5,6/m6} {
  \node at ([yshift=-1.1cm, xshift=-0.05cm]\x.west) {\i};
}

\node[stage, minimum width=0.5cm, fill=myblueRB, draw=myblueborderRB, thick, below=3.2cm of m1] (d1) {\textcolor{darkgray}{\shortstack{Load \\ 1}}};
\node[stage, minimum width=0.5cm, fill=myblueRB, draw=myblueborderRB, thick, right=of d1] (d2) {\textcolor{darkgray}{\shortstack{Load \\ 2}}};
\node[stage, fill=myblueRB, draw=myblueborderRB, thick, right=of d2] (d2a) {\textcolor{darkgray}{\shortstack{Load \\ 3}}};
\node[stage, minimum width=0.5cm, fill=mypinkRB, draw=mypinkborderRB, thick, right=of d2a] (d3) {\textcolor{darkgray}{\shortstack{Store \\ 1}}};
\node[stage, fill=mypinkRB, draw=mypinkborderRB, thick, right=of d3] (d4) {\textcolor{darkgray}{\shortstack{Store \\ 2}}};

\node[stage, minimum width=2.3cm, fill=mygreenRB!50, draw=mygreenborderRB!70, thick, above=0.1cm of d2, xshift=0.73cm] (d5) {\textcolor{darkgray}{\shortstack{Compute \\ 1}}};
\node[stage, fill=mygreenRB!50, draw=mygreenborderRB!70, thick, right=of d5, xshift=0.17cm] (d6) {\textcolor{darkgray}{\shortstack{Compute \\ 2}}};

\node[align=center, yshift=-2.8cm] at (l1.south) {\textbf{Decoupled access-execute}};

\draw[->, thick] 
  ([yshift=-0.6cm]d1.west) -- ([yshift=-0.6cm]d4.east);

\foreach \x in {d1, d2, d3} {
  \draw[dashed, thick, gray] 
    ([yshift=-0.8cm, xshift=-0.05cm]\x.west) -- ([yshift=1.70cm, xshift=-0.05cm]\x.west);
}

\draw[thick] 
  ([yshift=-0.7cm, xshift=-0.05cm]d1.west) -- ++(0,0.2);

\node[left = of d1, xshift=-0.1cm, yshift=-0.6cm] {\textbf{T}};

\foreach \i/\x in {0/d1,1/d2,2/d3} {
  \node at ([yshift=-1.1cm, xshift=-0.05cm]\x.west) {\i};
}

\draw[<->, thick, dashed, darkgray]
  ([yshift=-0.6cm]d4.east) -- ++(4.2,0) node[midway, above] {\textcolor{darkgray}{Latency gain}};

\end{tikzpicture}
    \caption{Comparison between a monolithic and chosen \ac{DAE} pipeline. In each discrete timestep, multiple datamover jobs can overlap with the computation.}
    \label{fig:DAE}
\end{figure}
improving resource utilization by memory latency hiding. As illustrated in~\cref{fig:DAE}, this is achieved by allowing compute and datamover jobs to run concurrently. 
\highlightRIII{
We use a discretized, software-managed time reference, called a tick, for scheduling, implemented in software on the controller core. Each timestep can host at most one compute job and an arbitrary number of data-transfer jobs. Data-transfer jobs assigned to the same timestep are executed sequentially by the single data-mover channel, while their execution can overlap with the compute job. Assuming a valid computation order, computing a tile requires at most three timesteps: one to push previous results from \ac{TCM} to off-chip memory, one to fetch input dependencies, and one to perform the computation. The push and fetch timesteps can be omitted when the corresponding data can be retained in \ac{TCM}. This model provides an upper bound on the number of timesteps required to execute the entire \ac{NN} and defines each tile's lifespan, thereby limiting the number of variables in the scheduling model.
}
\begin{figure}
\centering
    \begin{tikzpicture}[>=stealth, node distance=3cm, on grid, auto] 
    
  \node[state, fill=mygrayRB!50, draw=mygrayborderRB, very thick, minimum height=1.3cm, dashed, pattern=north east lines, pattern color=mygrayRB!70] (NE) { \textcolor{darkgray}{N/E}};
  \node[state, right=of NE, fill=mygrayRB!50, draw=mygrayborderRB, very thick, minimum height=1.3cm] (TCM) { \textcolor{darkgray}{TCM}};
  \node[state, below=of TCM, fill=mygrayRB!50, draw=mygrayborderRB, very thick, minimum height=1.3cm] (LCM) { \textcolor{darkgray}{DRAM}};
  \node[state, right=of TCM, fill=mygrayRB!50, draw=mygrayborderRB, very thick, minimum height=1.3cm] (T2CM) { \textcolor{darkgray}{L-TCM}};

  \path[->, myredborderRB, very thick] (NE) edge node { \textcolor{darkgray}{compute}} (TCM);
  \path[->, mygreenborderRB, very thick] (TCM) edge[bend left] node { \textcolor{darkgray}{push}} (LCM);
  \path[->, mygreenborderRB, very thick] (LCM) edge[bend left] node { \textcolor{darkgray}{fetch}} (TCM);
  \path[->, mygreenborderRB, very thick] (TCM) edge node { \textcolor{darkgray}{l-copy}} (T2CM);
  \path[->, mygreenborderRB, very thick] (LCM) edge[bend right] node[sloped, above] { \textcolor{darkgray}{l-fetch}} (T2CM);

  \path[->, thick, darkgray, dashed] (TCM) edge[bend left=20] (NE);
  \path[->, thick, darkgray, dashed] (LCM) edge[bend left=20] (NE);
  \path[->, thick, darkgray, dashed] (T2CM) edge[bend right=40] (NE);

\end{tikzpicture}
    \caption{Tile state diagram. Green arrows indicate transitions performed by the datamover, while the red arrow represents a transition executed by the compute unit.}
    \label{fig:statediagram}
\end{figure}
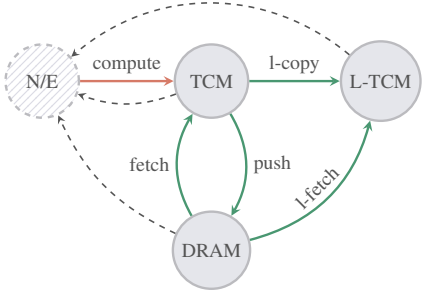

\textbf{Variables:} In the formulation, tiles are the atomic data blocks, and both operators and data-transfer jobs depend on them. Each tile is modeled according to the state diagram in~\cref{fig:statediagram}, which outlines four possible states for each tile:
\begin{itemize}
    \item \textit{N/E} $\Rightarrow$ outside its lifetime (i.e., not yet generated or no longer needed).
    \item \textit{DRAM} $\Rightarrow$  stored in off-chip memory.
    \item \textit{TCM} $\Rightarrow$  stored in \ac{TCM}.
    \item \textit{L-TCM} $\Rightarrow$ stored in \ac{TCM} and expanded to support line-parallel format.
\end{itemize}
All tiles initially start in the \textit{N/E} state, except for parameters and model inputs which start in \textit{DRAM}. State transitions are depicted by directed arrows; most are managed by the data mover, except for the transition from \textit{N/E} to \textit{TCM}, performed by the compute units. The available transitions are:
\begin{itemize}
    \item fetch $\Rightarrow$ copy tile from DRAM to TCM.
    \item push $\Rightarrow$ copy tile from TCM to DRAM.
    \item compute $\Rightarrow$ compute tile (writes to TCM).
    \item l-copy $\Rightarrow$ copies a tile within TCM to expand it into the line-parallel format.
    \item l-fetch $\Rightarrow$ fetches a tile into TCM in line-parallel format directly from DRAM. 
\end{itemize}

While remaining in a state incurs no cost, transitions introduce latency as moving or processing data consumes clock cycles. For completeness, special transitions are represented by dashed arrows, indicating that tile removal occurs without any associated cost, as used memory space can be overwritten. Transitions and states are both encoded as boolean variables, and each tile is associated with a set of these variables for every timestep it remains alive.

\textbf{Constraints:} Constraints ensure a valid scheduling behavior, and can be logically classified as follows.
    \paragraph{Persistency and Transition Constraints} States must persist across consecutive timesteps unless a transition occurs. For a tile \( j \) at timestep \( t \), the following constraint ensures that it can only reside in TCM if it was already there at the previous timestep or is being computed or fetched at the current timestep:
    \begin{align}
        &\text{TCM}(j, t - 1) 
        +\, \text{compute}(j, t-1) \notag \\
        &\quad +\, \text{fetch}(j, t-1)
        \geq\, \text{TCM}(j, t)\text{.}
    \label{persistency_constraint}
    \end{align}
    Similar constraints are present for the other states and transitions.
    
    \paragraph{Dependency Constraints} A tile can only be computed if all its input dependencies are present in \ac{TCM}. For tile \( j \) and its set of dependencies \( \text{dep}(j) \), this condition is enforced by:
    \begin{align}
    \forall j' \in \text{dep}(j), \quad \text{compute}(j, t) \leq \text{TCM}(j', t)\text{.}
    \end{align}
    Additionally, symmetric constraints are imposed when the computation requires input tiles to be in a line-parallel format.
        
    \paragraph{Bus Constraints} Shared buses between the datamover and the compute unit must not be used concurrently. 
    For a tile \( j \), computed at timestep \( t \), and sharing a bank with an adjacent tile \( j' \) of the same tensor, datamover operations on \( j' \) must be avoided:
    \begin{align}
        &\text{l-fetch}(j', t) 
        +\, \text{l-copy}(j', t) \notag \\
        &+\, \text{fetch}(j', t) 
        +\, \text{push}(j', t) 
        = 0\text{.}
    \label{bus_constraint}
    \end{align}
    Similar constraints are imposed for tiles that share a bank and are inputs to a computation.
    
    \paragraph{Memory Constraints} The total memory consumed by tiles in TCM must not exceed the available \ac{TCM}. At this stage, each tensor has already been  tiled, which means that the bank occupation of each tile is known. For each alive tensor \( i \in \mathcal{I}\) at timestep \( t \), let
    \[
    a_{i,j,t} =
    \begin{cases}
    1, & \text{if tile \( j \) of tensor \( i \) is active at time \( t \),} \\
    0, & \text{otherwise.}
    \end{cases}
    \]
    Then, it is possible to define two integer variables that represent the lowest and highest memory bank used per tensor $i$ at timestep $t$, based on the set of its active tiles:
    \begin{align}
    m_{i,t} = \min \{ L(j, t) \mid a_{i,j,t} = 1 \}. \\ M_{i,t} = \max \{ H(j, t) \mid a_{i,j,t} = 1 \}.
    \end{align}
    where $H(j, t)$ and $L(j, t)$ indicate the highest and lowest bank used by the tile $j$ at timestep $t$.
    The inclusion of the timestep in these expressions accounts for memory reuse: under certain conditions, output tiles are allowed to overwrite input tiles, thereby reducing the combined memory footprint. These reuse configurations, specifying which tiles can be overwritten and how many banks are reused, are determined prior to the scheduling phase.
    Then, the memory occupied by tensor \( i \) at time \( t \), independently of where it will be effectively stored in memory, is given by:
    \begin{align}
    \text{Mem}_{i,t} = M_{i,t} - m_{i,t} + 1.
    \end{align}
    For each timestep, the total memory occupation over all active tensors must satisfy the constraint
    \begin{align}
        \sum_{i \in \mathcal{I}} \text{Mem}_{i,t} = \sum_{i \in \mathcal{I}} M_{i,t} - m_{i,t} + 1 \le C.
        \label{memory_constraint}
    \end{align}
    where \( C \) is the number of TCM banks available.

These constraints collectively define the feasible region for valid scheduling.
However, the goal is not just to find any valid schedule but to optimize it for minimal latency. This is achieved by defining an objective function.

\textbf{Objective Function:} The objective function to be minimized represents the total execution latency, computed as the sum of the latencies across all considered timesteps:
\begin{equation}
    min \quad \delta N_{DM} + \sum_{t=0}^{\text{T}} max(l_{DM}(t), l_{C}(t))\text{,}
\end{equation}
where $l_{DM}(t)$ and $l_{C}(t)$ denote the latencies of the datamover and compute unit at timestep $t$, respectively, and $N_{DM}$ represents the total number of datamover operations. Since the datamover and compute unit operate in parallel, the latency for a single timestep is determined by the maximum of the two. The additional term $\delta N_{DM}$ introduces a tuneable penalty to discourage unnecessary datamover operations that are hidden behind computation while still requiring system bus and DDR bandwidth potentially affecting the performance of other components in the SoC\@.
The \ac{CP} solver then optimizes the schedule, retaining only the necessary timesteps and eliminating those without any transitions.

\textbf{Scalability:} 
As each tile is associated with a set of variables for every timestep it remains active, the total number of variables grows linearly with both the number of timesteps and tiles. However, given our time modelization, the total number of timesteps is exactly three times the number of tiles, causing the scheduling problem size to scale quadratically with the number of tiles. As a result, building and solving a single scheduling problem does not scale well with the dimension of the \ac{NN}\@. Instead, breaking down the monolithic problem into smaller subproblems significantly improves compilation times. It introduces only a minor performance trade-off, due to a reduced ability to overlap jobs, as the solver no longer has a complete view of the entire scheduling space.~\cref{tab:scalability} shows the impact of this trade-off on a real-case scenario.

\subsection{Optimizations} \label{subsec:opt}
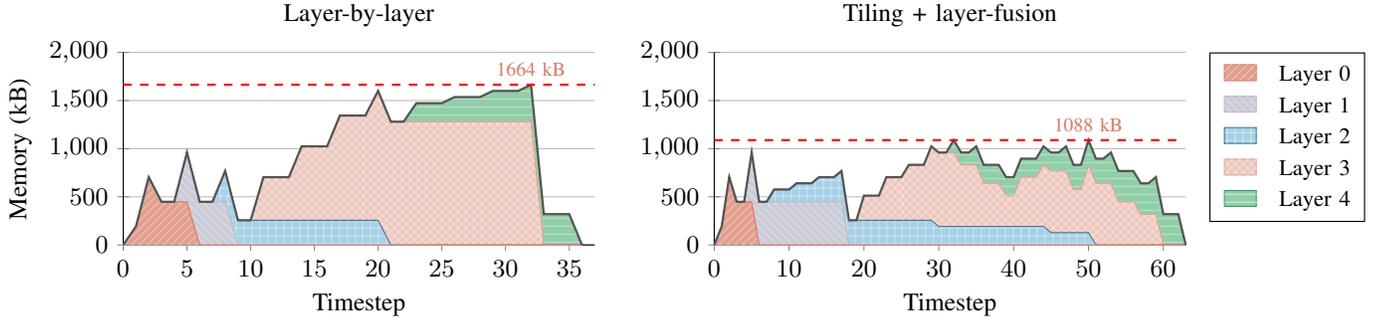
\begin{figure*}[t]
\centering
    \begin{tikzpicture}
\small
\begin{groupplot}[
    group style={
        group size=2 by 1,
        horizontal sep=1.5cm
    },
    width=7.5cm,
    height=4cm,
    xlabel={Timestep},
    ylabel={Memory (kB)},
    ymin=0,
    ymax=2000,
    xtick distance=5,
    ymajorgrids=true,
    axis lines=left,
    enlarge x limits=false,
    cycle list name=black white,
]
\nextgroupplot[
    title={Layer-by-layer},
    legend style={draw=none},    
    axis line style={-}
]

\pgfplotstableread[col sep=comma]{alive_memory_first.csv}\originaldata

\addplot[draw=none, fill=mygreenRB] table[
    x=timestep,
    y expr=4*(\thisrow{layer_0} + \thisrow{layer_1} + \thisrow{layer_2} + \thisrow{layer_3} + \thisrow{layer_4})
] {\originaldata};

\addplot[draw=none, pattern=horizontal lines, pattern color=white, opacity=0.6] table[
    x=timestep,
    y expr=4*(\thisrow{layer_0} + \thisrow{layer_1} + \thisrow{layer_2} + \thisrow{layer_3} + \thisrow{layer_4})
] {\originaldata};

\addplot[draw=none, fill=mypinkRB] table[
    x=timestep,
    y expr=4*(\thisrow{layer_0} + \thisrow{layer_1} + \thisrow{layer_2} + \thisrow{layer_3})
] {\originaldata};

\addplot[draw=none, pattern=crosshatch, pattern color=white, opacity=0.6] table[
    x=timestep,
    y expr=4*(\thisrow{layer_0} + \thisrow{layer_1} + \thisrow{layer_2} + \thisrow{layer_3})
] {\originaldata};

\addplot[draw=mypinkborderRB, thin] table[
    x=timestep,
    y expr=4*(\thisrow{layer_0} + \thisrow{layer_1} + \thisrow{layer_2} + \thisrow{layer_3})
] {\originaldata};

\addplot[draw=none, fill=myblueRB] table[
    x=timestep,
    y expr=4*(\thisrow{layer_0} + \thisrow{layer_1} + \thisrow{layer_2})
] {\originaldata};

\addplot[draw=none, pattern=grid, pattern color=white, opacity=0.6] table[
    x=timestep,
    y expr=4*(\thisrow{layer_0} + \thisrow{layer_1} + \thisrow{layer_2})
] {\originaldata};

\addplot[draw=myblueborderRB, thin] table[
    x=timestep,
    y expr=4*(\thisrow{layer_0} + \thisrow{layer_1} + \thisrow{layer_2})
] {\originaldata};

\addplot[draw=none, fill=mygrayRB] table[
    x=timestep,
    y expr=4*(\thisrow{layer_0} + \thisrow{layer_1})
] {\originaldata};

\addplot[draw=none, pattern=north west lines, pattern color=white, opacity=0.6] table[
    x=timestep,
    y expr=4*(\thisrow{layer_0} + \thisrow{layer_1})
] {\originaldata};

\addplot[draw=mygrayborderRB, thin] table[
    x=timestep,
    y expr=4*(\thisrow{layer_0} + \thisrow{layer_1})
] {\originaldata};

\addplot[draw=none, fill=myredRB] table[
    x=timestep,
    y expr=4*\thisrow{layer_0}
] {\originaldata};

\addplot[draw=none, pattern=north east lines, pattern color=white, opacity=0.6] table[
    x=timestep,
    y expr=4*\thisrow{layer_0}
] {\originaldata};

\addplot[draw=myredborderRB, thin] table[
    x=timestep,
    y expr=4*\thisrow{layer_0}
] {\originaldata};

\addplot+[darkgray, thick, mark=none] table[x=timestep, y expr=4*\thisrow{total}] {\originaldata};

\addplot[red, dashed, thick] coordinates {
    (0,1664)(37,1664)
};
\node[anchor=south, text=myredborderRB] at (axis cs:32,1664) {\scriptsize 1664 kB};

\nextgroupplot[
    title={Tiling + layer-fusion},
    ylabel={},  
    axis line style={-},
    legend style={at={(1.05,1)}, anchor=north west, font=\footnotesize},
    xtick distance=10,
]

\addlegendimage{empty legend}
\addlegendentry{
    \tikz[baseline={(0,0)}]{
        \path[fill=myredRB, draw=myredborderRB] (0,0) rectangle (0.4cm,0.2cm);
        \path[pattern=north east lines, pattern color=white, opacity=0.6] (0,0) rectangle (0.4cm,0.2cm);
    }
    \hspace{0.3em} Layer 0
}

\addlegendimage{empty legend}
\addlegendentry{
    \tikz[baseline={(0,0)}]{
        \path[fill=mygrayRB, draw=mygrayborderRB] (0,0) rectangle (0.4cm,0.2cm);
        \path[pattern=north west lines, pattern color=white, opacity=0.6] (0,0) rectangle (0.4cm,0.2cm);
    }
    \hspace{0.3em} Layer 1
}

\addlegendimage{empty legend}
\addlegendentry{
    \tikz[baseline={(0,0)}]{
        \path[fill=myblueRB, draw=myblueborderRB] (0,0) rectangle (0.4cm,0.2cm);
        \path[pattern=grid, pattern color=white, opacity=0.6] (0,0) rectangle (0.4cm,0.2cm);
    }
    \hspace{0.3em} Layer 2
}

\addlegendimage{empty legend}
\addlegendentry{
    \tikz[baseline={(0,0)}]{
        \path[fill=mypinkRB, draw=mypinkborderRB] (0,0) rectangle (0.4cm,0.2cm);
        \path[pattern=crosshatch, pattern color=white, opacity=0.6] (0,0) rectangle (0.4cm,0.2cm);
    }
    \hspace{0.3em} Layer 3
}

\addlegendimage{empty legend}
\addlegendentry{
    \tikz[baseline={(0,0)}]{
        \path[fill=mygreenRB, draw=mygreenborderRB] (0,0) rectangle (0.4cm,0.2cm);
        \path[pattern=horizontal lines, pattern color=white, opacity=0.6] (0,0) rectangle (0.4cm,0.2cm);
    }
    \hspace{0.3em} Layer 4
}

\pgfplotstableread[col sep=comma]{alive_memory_second.csv}\optimizeddata
    
\addplot[draw=none, fill=mygreenRB] table[
    x=timestep,
    y expr=4*(\thisrow{layer_0} + \thisrow{layer_1} + \thisrow{layer_2} + \thisrow{layer_3} + \thisrow{layer_4})
] {\optimizeddata};

\addplot[draw=none, pattern=horizontal lines, pattern color=white, opacity=0.6] table[
    x=timestep,
    y expr=4*(\thisrow{layer_0} + \thisrow{layer_1} + \thisrow{layer_2} + \thisrow{layer_3} + \thisrow{layer_4})
] {\optimizeddata};

\addplot[draw=none, fill=mypinkRB] table[
    x=timestep,
    y expr=4*(\thisrow{layer_0} + \thisrow{layer_1} + \thisrow{layer_2} + \thisrow{layer_3})
] {\optimizeddata};

\addplot[draw=none, pattern=crosshatch, pattern color=white, opacity=0.6] table[
    x=timestep,
    y expr=4*(\thisrow{layer_0} + \thisrow{layer_1} + \thisrow{layer_2} + \thisrow{layer_3})
] {\optimizeddata};

\addplot[draw=mypinkborderRB, thin] table[
    x=timestep,
    y expr=4*(\thisrow{layer_0} + \thisrow{layer_1} + \thisrow{layer_2} + \thisrow{layer_3})
] {\optimizeddata};

\addplot[draw=none, fill=myblueRB] table[
    x=timestep,
    y expr=4*(\thisrow{layer_0} + \thisrow{layer_1} + \thisrow{layer_2})
] {\optimizeddata};

\addplot[draw=none, pattern=grid, pattern color=white, opacity=0.6] table[
    x=timestep,
    y expr=4*(\thisrow{layer_0} + \thisrow{layer_1} + \thisrow{layer_2})
] {\optimizeddata};

\addplot[draw=myblueborderRB, thin] table[
    x=timestep,
    y expr=4*(\thisrow{layer_0} + \thisrow{layer_1} + \thisrow{layer_2})
] {\optimizeddata};

\addplot[draw=none, fill=mygrayRB] table[
    x=timestep,
    y expr=4*(\thisrow{layer_0} + \thisrow{layer_1})
] {\optimizeddata};

\addplot[draw=none, pattern=north west lines, pattern color=white, opacity=0.6] table[
    x=timestep,
    y expr=4*(\thisrow{layer_0} + \thisrow{layer_1})
] {\optimizeddata};

\addplot[draw=mygrayborderRB, thin] table[
    x=timestep,
    y expr=4*(\thisrow{layer_0} + \thisrow{layer_1})
] {\optimizeddata};

\addplot[draw=none, fill=myredRB] table[
    x=timestep,
    y expr=4*\thisrow{layer_0}
] {\optimizeddata};

\addplot[draw=none, pattern=north east lines, pattern color=white, opacity=0.6] table[
    x=timestep,
    y expr=4*\thisrow{layer_0}
] {\optimizeddata};

\addplot[draw=myredborderRB, thin] table[
    x=timestep,
    y expr=4*\thisrow{layer_0}
] {\optimizeddata};

\addplot+[darkgray, solid, thick, mark=none] table[x=timestep, y expr=4*\thisrow{total}] {\optimizeddata};

\addplot[red, dashed, thick] coordinates {
    (0,1088)(62,1088)
};
\node[anchor=south, text=myredborderRB] at (axis cs:50,1088) {\scriptsize 1088 kB};

\end{groupplot}
\end{tikzpicture}
    \caption{Memory usage over time during the execution of the first five layers of MobileNetV2, with and without the proposed optimization for layer fusion and tiling. Colored/patterned regions correspond to the memory footprint of individual layers.} 
    \label{fig:memoryusage}
\end{figure*}

Temporal tiling is a key aspect of the compilation process. Larger tiles may consume excessive \ac{TCM} space, leaving insufficient memory for buffering and thus limiting the scheduler's ability to hide memory latency, while overly small tiles may significantly increase compilation time.
While most existing approaches emphasize micro-tiling and overlook compilation time as a major factor~\cite{klotski, stream-dse}, our approach seeks a balanced compromise between performance and compilation overhead, avoiding excessive trade-offs on either side. 
Additionally, tile computation order also directly affects latency. In conventional layer-by-layer execution, tiles from different layers cannot be interleaved, requiring all tiles of a layer to be processed consecutively. However, interleaving layer execution, commonly known as layer fusion, can reduce overall off-chip memory usage and thus improve execution latency. As shown in~\cref{fig:memoryusage}, properly combining layer fusion and tiling can significantly reduce memory requirements in real-world scenarios.

\textbf{Problem Formulation:}
The goal is to optimize global memory usage by selecting appropriate tile sizes for each tensor while considering layer fusion. The optimization model used is a simplified variant of the one described in \cref{subsec:scheduling}. Unlike the scheduling problem, however, this formulation is not constrained by the actual on-chip memory capacity. Instead, it focuses on minimizing the volume of data that must be offloaded to off-chip memory during scheduling. To this end, the model simplifies the memory hierarchy by considering only a single level, targeting the reduction of total memory usage at each step.

\textbf{Variables:} 
To determine the optimal tile sizes, dependencies between tensors are ideally analyzed line by line, like in micro-tiling strategies. However, as previously noted, micro-tiling impairs the runtime for complex networks dramatically. To address this, our model simplifies the search space by limiting tile size choices: the largest tile that fits within \ac{TCM}, and tile sizes reduced by fixed factors.
For each feature map $i$, tiles for each considered size are generated. The selection of the tile size for each data tensor is represented by one boolean variable per size option:
$LS_{k,i}$, indicating the selection of the $k^\text{th}$ biggest tile size for tensor $i$. Only the tiles of the selected size will allocate on-chip memory. To optimize compiler runtime, in the remainder of this work we consider only two tile-size options per layer. 

In this formulation, there is only one memory hierarchy, so for each tile, there are two states:
\begin{itemize}
    \item \textit{N/E} $\Rightarrow$ the tile is outside its lifetime.
    \item \textit{TCM} $\Rightarrow$ the tile is stored in TCM.
\end{itemize}
All tiles initially start in the N/E state, except for parameters and model inputs that are already assumed in TCM\@. \text{Compute} operations move tile states from N/E to TCM.

Because the optimization target is to reduce the on-chip memory overhead, we introduce another integer variable: $MemTh_{t}$, indicating the memory needed at timestep $t$ of the execution, considering all the tiles in TCM.

\textbf{Constraints:} 
The constraint expressed in \cref{memory_constraint} is modified to take into account the new variable:

\begin{align}
    \sum_{i \in \mathcal{I}} \text{Mem}_{i,t} \le \text{MemTh}_{t}.
    \label{memory_constraint_fusion}
\end{align}
The memory consumption per timestep is not constrained to a fixed value, and the variable $MemTh_{t}$ is used to track it.
Because there is one memory hierarchy, this formulation does not consider the Bus constraints \#\ref{bus_constraint}. Also, the Persistency and Transition constraints \#\ref{persistency_constraint} are modified, removing the $fetch$ transition.

Only one tile size is selected for each tensor:
\begin{align}
    \forall i \in \text{tensors}, \quad \sum_k\text{LS}_{k,i} = 1\text{.}
\end{align}

Lastly, only tiles related to the selected tile size $LS_{k,i}$ are enabled:
\begin{align}
    \forall j \in tiles, \quad \text{TCM}(j, t) \leq \text{LS}_{k,i}\text{.}
\end{align}

\textbf{Objective Function:} 
\highlightRIII{
The objective function minimizes the aggregate memory demand exceeding the available \ac{TCM} capacity $C$ across all timesteps:
\begin{equation}
    \min \quad \sum_{t=0}^{T} \max \left(0,\,\left(MemTh_t - C\right)\right)\text{.}
\end{equation}
}
\textbf{Scalability:}  
The proposed problem introduces a simplified model that reduces the number of states by half compared to the scheduling formulation and decreases the number of timesteps by a factor of three, as it does not account for additional memory hierarchies. The problem is further decomposed by identifying regions where activation data cannot be held entirely on-chip and restricting layer fusion only to those areas. How this division impacts inference time is evaluated in~\cref{tab:scalability}.

\begin{table}[t]
     \footnotesize
     \centering
     \caption{Impact of problem partitioning on YOLOv8N-det compilation and inference times.}
     \label{tab:scalability}
     \begin{tabular}{
        l
        S[table-format=4.0, table-space-text-post={\enspace(\%)}]
        S[table-format=2.2, table-space-text-post={\enspace(\%)}]
        }
        \toprule
        \textbf{Problem partitioning} & \textbf{Compilation Time (s)} & \textbf{Inference Time (ms)} \\
        \midrule
        No partitioning       & 3480 \hspace{1mm} (0.0\%)    & 23.9 \hspace{1mm} (0.0\%)    \\
        Only optimizations       & 1331 \hspace{1mm} (-61.7\%)  & 24.0 \hspace{1mm} (+0.5\%)   \\
        Only scheduling     & 1423 \hspace{1mm} (-59.1\%)  & 24.5 \hspace{1mm} (+2.4\%)   \\
        Both            & 667  \hspace{1mm} (-80.8\%)  & 24.7 \hspace{1mm} (+3.3\%)   \\
        \bottomrule
     \end{tabular}
\end{table}

\subsection{Memory Allocation}
As described in~\cref{sec:npu_arch}, the virtual mapping of the physical space can be changed through the \ac{V2P} operation.
Given the sequence of timed jobs, the allocation step assigns virtual memory addresses by reserving space for each tile in the virtual view of the \ac{TCM}. Alongside this, it determines the corresponding physical bank mappings and generates the necessary \ac{V2P} table updates, ensuring that the compute engine perceives a contiguous view of its data. The constraints defined in the scheduling problem ensure that the \ac{TCM} space is enough to store all the active tiles in each timestep, guaranteeing that a feasible allocation always exists. However, the allocation step must also account for other architectural choices: 

    \paragraph{Virtual space contiguity} Tiles of the same spatial-tile of a tensor must be stored sequentially in virtual memory. Otherwise, consumers of the tensor may not be able to read it correctly, as the receptive field could span multiple input tiles.
    \paragraph{Physical space preservation} Tiles must maintain their physical memory banks across timesteps.
    \paragraph{Reuse optimization} Output tensors must be  placed before input tensors in virtual memory, respecting the correct distance, to enable output-input overwriting. This allows consumed data to be overwritten, reducing overall memory usage.
    \paragraph{Bank exclusivity} Different tensors, used in the same timestep,  cannot share the same bank to prevent bus conflicts.

To ensure all these properties are met, the allocation step is formalized as a \ac{CP} problem, similar to scheduling. Moreover, like the scheduling phase, the allocation problem can be decomposed into smaller subproblems that are solved independently, improving scalability and efficiency.

\section{Experimental Results}
\label{sec:results}
\begin{table*}[htbp]
\scriptsize
\centering
\setlength{\tabcolsep}{3.32pt}
\caption{\highlightRI{End-to-end model inference results on computer-vision models.
For each NPU, peak TOPS, DRAM bandwidth, and SRAM size are reported together with batch-size-1 end-to-end inference latency and LTP.
The best value per metric is highlighted.}
}
\label{tab:results}
\begin{tabular}{l|c|lcc|lcc|lcc|lcc}
\toprule
\multirow{3}{*}{\textbf{NPU}} &
\multirow{3}{*}{\shortstack{\textbf{HW Config.}\\\textbf{TOPS / DDR-BW / SRAM}}} &
\multirow{3}{*}{\textbf{Model}} &
\multicolumn{2}{c|}{\textbf{Results} $\downarrow$} &
\multirow{3}{*}{\textbf{Model}} &
\multicolumn{2}{c|}{\textbf{Results  $\downarrow$}} &
\multirow{3}{*}{\textbf{Model}} &
\multicolumn{2}{c|}{\textbf{Results  $\downarrow$}} &
\multirow{3}{*}{\textbf{Model}} &
\multicolumn{2}{c}{\textbf{Results  $\downarrow$}} \\

& & &
\textbf{Lat.} & \textbf{LTP} &
& \textbf{Lat.} & \textbf{LTP} &
& \textbf{Lat.} & \textbf{LTP} &
& \textbf{Lat.} & \textbf{LTP} \\

& & &
\textbf{[ms]} & &
& \textbf{[ms]} & &
& \textbf{[ms]} & &
& \textbf{[ms]} & \\
\midrule
\midrule

Ours$^M$
& 2 / 12\,GB/s / 1\,MB
& & 1.0 & \underline{\textbf{2.1}}
& & 0.9 & \underline{\textbf{1.8}}
& & 0.8 & \underline{\textbf{1.6}}
& & 7.0 & \underline{\textbf{14.0}} \\

\ac{eNPU}-A$^V$
& 2 / 12\,GB/s / 1\,MB
& MobileNet & 1.3 & 2.6
& MobileNet & 1.5 & 3.1
& MobileNet & 1.1 & 2.1
& ResNet & 8.5 & 16.9 \\

\ac{eNPU}-B$^V$
& 4 / 24\,GB/s / 2\,MB
& V1 & 0.8 & 3.2
& V2 & 1.0 & 4.0
& V3 & \underline{\textbf{0.7}} & 2.9
& 50V1 & 5.5 & 22.0 \\

\ac{iNPU}$^D$
& 11 / -- / --
& & \underline{\textbf{$\geq$0.3}} & $\geq$3.9
& & \underline{\textbf{$\geq$0.3}} & $\geq$3.2
& & $\geq$2.0 & $\geq$22.3
& & \underline{\textbf{$\geq$4.5}} & $\geq$49.4 \\

\midrule

Ours$^M$
& 2 / 12\,GB/s / 1\,MB
& & 1.3 & \underline{\textbf{2.7}}
& & 3.6 & \underline{\textbf{7.3}}
& & 24.6 & 49.2
& & 42.5 & \underline{\textbf{85.0}} \\

\ac{eNPU}-A$^V$
& 2 / 12\,GB/s / 1\,MB
& EfficientNet & 1.7 & 3.3
& EfficientDet & 5.3 & 10.6
& YOLOv8 & 98.2 & 196.5
& YOLOv8 & 115.8 & 231.6 \\

\ac{eNPU}-B$^V$
& 4 / 24\,GB/s / 2\,MB
& Lite0 & \underline{\textbf{1.0}} & 4.2
& Lite0 & \underline{\textbf{3.1}} & 12.5
& N-det. & 81.9 & 327.5
& S-det. & 91.4 & 365.8 \\

\ac{iNPU}$^D$
& 11 / -- / --
& & $\geq$3.3 & $\geq$36.5
& & $\geq$8.7 & $\geq$95.1
& & \underline{\textbf{$\geq$3.5}} & \underline{\textbf{$\geq$38.8}}
& & \underline{\textbf{$\geq$10.6}} & $\geq$116.9 \\

\midrule

Ours$^M$
& 2 / 12\,GB/s / 1\,MB
& & 28.7 & \underline{\textbf{57.4}}
& & 7.2 & \underline{\textbf{14.4}}
& & 2.6 & \underline{\textbf{5.2}}
& & 2.7 & \underline{\textbf{5.4}} \\

\ac{eNPU}-A$^V$
& 2 / 12\,GB/s / 1\,MB
& YOLOv8 & 101.8 & 203.5
& DAMO & 8.0 & 15.9
& MobileNet & 2.7 & 5.4
& MobileNet & 3.2 & 6.4 \\

\ac{eNPU}-B$^V$
& 4 / 24\,GB/s / 2\,MB
& N-seg. & 84.0 & 335.8
& YOLO-NL & \underline{\textbf{4.0}} & 16.1
& V1 SSD & \underline{\textbf{1.9}} & 7.5
& V2 SSD & \underline{\textbf{2.2}} & 8.8 \\

\ac{iNPU}$^D$
& 11 / -- / --
& & \underline{\textbf{$\geq$12.9}} & $\geq$141.5
& & $\geq$9.0 & $\geq$98.6
& & $\geq$4.5 & $\geq$49.2
& & $\geq$6.7 & $\geq$73.4 \\

\bottomrule
\multicolumn{14}{l}{\textbf{Superscript Legend:} \quad
$M$: Measured results \quad
$V$: Vendor reported \quad
$D$: Derived from vendor reported data (best-case assumption)}
\end{tabular}
\end{table*}

\highlightRIII{
We evaluate the proposed hardware--software system on a broad set of standard computer-vision models covering image classification, object detection, and semantic segmentation. All evaluated models use INT8 activations and weights and are exported using LiteRT v2.18.0. The models are benchmarked as provided, without additional approximations such as weight sparsification or further quantization. Consequently, the proposed deployment and optimization flow does not modify the quantized model parameters or introduce additional model-level approximations that would affect the reference model accuracy. We therefore report the reference accuracy of the evaluated models but focus the following evaluation on end-to-end inference performance.
Image-classification models target the ImageNet dataset~\cite{deng2009imagenet}, while object-detection and semantic-segmentation models target COCO 2017~\cite{lin2014coco}.
} A list of models and their characteristics is provided in \cref{tab:models}.

\begin{table}[ht]
\centering
\scriptsize
\caption{\highlightRIII{Models used for validation. Size denotes the number of parameters.
Accuracy denotes ImageNet top-1 accuracy or COCO AP ($\mathrm{AP}_{50:95}$) for INT8 quantized models.}}
\label{tab:models}
\begin{tabular}{lccc}

\toprule

\multicolumn{4}{c}{\textbf{ImageNet Classification ($224\times 224$ input)}} \\

\midrule
\textbf{Model} & \textbf{MACs [G]} & \textbf{Size [M]} & \textbf{Acc. [\%]} \\
\midrule
MobileNetV1            & 0.57 &  4.2 & 69.6 \\
MobileNetV2            & 0.30 &  3.5 & 69.8 \\
MobileNetV3-Large Min. & 0.21 &  3.9 & 71.3 \\
ResNet-50 V1           & 4.09 & 25.6 & 76.1 \\
EfficientNet-Lite0     & 0.41 &  4.7 & 74.4 \\
\bottomrule
\end{tabular}
\par\vspace{10pt}
\begin{tabular}{lcccc}
\toprule
\multicolumn{5}{c}{\textbf{COCO 2017 Detection / Segmentation}} \\
\midrule
\textbf{Model} & \textbf{Resolution} & \textbf{MACs [G]} &
\textbf{Params [M]} & \textbf{AP [\%]} \\
\midrule
EfficientDet-Lite0 & 320$\times$320 &  1.27 &  3.2 & 26.1 \\
YOLOv8n            & 640$\times$640 &  4.35 &  3.2 & 34.7 \\
YOLOv8s            & 640$\times$640 & 14.30 & 11.2 & 42.5 \\
YOLOv8n-seg        & 640$\times$640 &  6.30 &  3.4 & 28.8 \\
SSD MobileNetV1    & 300$\times$300 &  1.30 &  5.1 & 18.2 \\
SSD MobileNetV2    & 300$\times$300 &  0.80 &  4.3 & 22.0 \\
DAMO-YOLO-Nl       & 416$\times$416 &  3.02 &  5.7 & 40.2 \\
\bottomrule
\end{tabular}
\end{table}

\highlightRIII{Silicon performance is measured as batch-size-1 end-to-end inference latency on a production, mass-market \ac{MPU} integrating the proposed 2-TOPS \ac{NPU} at \SI{1}{GHz}, with 1\,MiB of SRAM and \SI{12}{GB/s} of DDR bandwidth. Measurements use the LiteRT benchmarking tools and unmodified INT8 per-channel-quantized Kaggle models~\cite{tensorflow_benchmark_tool}. Silicon measurements and compiler estimates differed by less than \SI{3}{\%} in all cases, showing the validity of the hardware models in the compiler.}

\highlightRIII{
The constraint models were solved using the CP-SAT solver in Google OR-Tools v9.12 on an Intel Xeon W-3335 system (16 cores, 512\,GiB main memory) running Ubuntu 22.04.5 LTS. CP-SAT was configured with 16 search workers, a 200\,s wall-clock time limit, and a 1\% relative optimality-gap limit. All other solver parameters were kept at their default values.
}

To compare our proposed technique with the state-of-the-art, we benchmark public, general-purpose embedded NPU (\acs{eNPU}) IPs that allow estimating real-life end-to-end latency on modern AI benchmarks. The results for the \acs{eNPU}~\cite{arm_ethos_u85} with the best overall peak performance and performance--cost trade-off are used in this paper in two configurations:
\highlightRIII{
\begin{enumerate}[label=\alph*)]
    \item \emph{\acs{eNPU}-A}: \SI{2}{TOPS}, \SI{1}{\GHz}, \SI{1}{\mebi\byte} SRAM (\SI{64}{GB/s}), \SI{12}{GB/s} DRAM.
    \item \emph{\acs{eNPU}-B}: \SI{4}{TOPS}, \SI{1}{\GHz}, \SI{2}{\mebi\byte} SRAM (\SI{128}{GB/s}), \SI{24}{GB/s} DRAM.
\end{enumerate}
}
Configuration A matches the system resources of our proposed NPU, while configuration B represents a higher-end system with 2$\times$ the resources in both DDR bandwidth and TCM capacity, aiming to illustrate that our approach can compete with state-of-the-art designs at half the provisioned TOPS\@.

\highlightRII{
In addition to the \acs{eNPU}s, we compare against benchmarkable AI-Vision-processor systems integrating an \acs{iNPU} below 20\,TOPS. For a strong comparison, we include the system in Ref.~\cite{hailo_15_ai_vision_processor}, which provides the best performance--resource trade-off on the considered benchmarks among the evaluated systems in this class. This \ac{iNPU} delivers up to 11\,TOPS and is optimized for computer-vision workloads. Its exact TCM capacity is not disclosed but can be derived from vendor-reported information to be in the range of tens of MiB, while its DDR bandwidth is approximately 17\,GB/s (32\,bit at 4266\,MT/s). Both resource provisions are therefore higher than those of our proposed NPU system.
}

\highlightRIII{
Performance data for the \acs{eNPU}s were obtained using the vendor's toolchain in version 4.4 with the same models evaluated on our proposed NPU system. Since area and power estimates are not available for the reference systems, the comparison primarily focuses on performance in terms of batch-size-1 end-to-end inference latency. For the resource comparison, peak TOPS and DDR bandwidth are used as proxies for compute-related area and memory-related power, respectively, reflecting the substantial contributions of MAC resources to accelerator area and DRAM data movement to system power~\cite{dally2024throughput}. In both proxy metrics, our proposed system is better than or on par with all reference systems.
}

\highlightRI{

The \acs{iNPU} performance is assessed thorough the published model zoo~\cite{hailo_model_explorer}. However, the model zoo reports throughput rather than end-to-end inference latency. Due to inference dataflow pipelining, throughput does not directly determine the latency of an individual inference. The latter depends on the pipeline depth, which is not publicly reported. In the absence of explicit latency data, we therefore approximate latency as the inverse of the reported throughput, corresponding to the optimistic assumption of a pipeline depth of one. This represents an asymptotic best-case latency estimate for the \acs{iNPU}; any additional pipeline depth would increase its actual inference latency. Thus, the approximation favors the reference system and provides a conservative basis for assessing the latency advantage of the proposed architecture.
}


\highlightRI{
To capture performance efficiency relative to provisioned compute, we introduce a second metric in addition to latency: the Latency--TOPS Product (LTP), defined as $\text{latency} \cdot \text{TOPS}$,
where lower values indicate higher efficiency. Since peak \ac{TOPS} is proportional to the deployed \ac{MAC}-unit count and clock frequency, it provides a proxy for provisioned compute resources under comparable technology, precision, and operating conditions~\cite{sze2020evaluate}. LTP therefore quantifies the peak compute provisioned to achieve a given end-to-end inference latency and, under these conditions, provides a cost--delay-oriented measure of compute efficiency. This distinction is motivated by the results in \cref{tab:npus_eTOPS}: two \acp{NPU} achieving equal latency are not equivalent in compute efficiency if one provisions hardware for up to 11\,TOPS while the other provisions only for up to \SI{4}{TOPS}.
}

The experimental results are summarized in \cref{tab:results}. Compared with the \acs{eNPU} with identical resources, our system achieves an average speedup of \SI{1.8}{\times} across all benchmarks, with gains up to \SI{4}{\times} (YOLOv8N detection). \highlightRIII{Here and in the following, the average speedup is computed as the arithmetic mean across all evaluated benchmarks of the per-benchmark speedup ratios, defined as the reference NPU latency divided by the achieved latency of our NPU}. Against the 2$\times$ larger \acs{eNPU}-B, our approach still delivers an average performance uplift of \SI{1.3}{\times}, peaking at \SI{3.3}{\times}. Improvements are most pronounced for complex/compute-intensive models such as YOLOv8, demonstrating the superior scalability of our compiler for advanced neural networks.

Even compared to the 11\,TOPS \acs{iNPU} system, our 2\,TOPS system achieves an average speedup of \SI{1.25}{\times} and up to \SI{2.5}{\times} on individual benchmarks—despite having more than $5\times$ fewer provisioned TOPS\@. Across all but one case, our design achieves the best LTP, confirming superior performance-per-TOPS under strict memory and bandwidth constraints. These findings confirm that intelligent HW/SW co-design, not raw TOPS, determines real-world edge-AI performance.

\subsection{Ablation Study}
\label{sec:ablation}

\highlightRI{
To quantify the contribution of the compiler optimizations to end-to-end inference latency, we perform a cumulative ablation study across all evaluated models. \cref{fig:ablation} reports the normalized latency overhead for four configurations (P0--P3), with their enabled features summarized in \cref{tab:config_features}. P0 enables all evaluated optimizations---Layer fusion, Hide \ac{DM}, and Both Parallelisms---and serves as the baseline corresponding to the latency results in \cref{tab:results}. P1 disables Layer fusion, P2 additionally disables Hide \ac{DM}, and P3 further disables the remaining parallelism optimization, retaining only Depth Parallelism. Across all evaluated models, progressively disabling these optimizations increases end-to-end inference latency, although the magnitude depends on the workload. The largest increases occur for several of the more compute-intensive models, demonstrating the importance of combining the compiler optimizations rather than relying on any single mechanism.
}

\begin{figure}[t]
\centering
\pgfplotstableread[col sep=comma]{barplot_data.csv}\bardata

\begin{tikzpicture}
\small
\begin{axis}[scale only axis,
    width=0.85\columnwidth,
    height=0.42\columnwidth,
    ybar stacked,
    bar width=6pt,
    symbolic x coords={
        V2 SSD, V1 SSD, YOLO-NL,
        YOLOv8 N-det., YOLOv8 N-seg., YOLOv8 S-det.,
        EfficientNet Lite0, MobileNet V2, MobileNet V3, MobileNet V1, ResNet 50V1, EfficientDet Lite0
    },
    xtick=data,
    x tick label style={
        rotate=45,
        anchor=east,
        font=\scriptsize
    },
    ylabel={Latency overhead},
    ylabel style={font=\footnotesize},
    ytick={1,2},
    yticklabels={1x, 2x},
    ymin=0,
    ymajorgrids=true,
    grid style={dashed, gray!40},
    axis lines=left,
    axis line style={-},
    enlarge x limits=0.06,
    legend style={
        at={(0.99,0.99)},
        anchor=north east,
        font=\footnotesize,
        draw=none,
        row sep=0.5pt,
        legend columns=2,
    },
    legend cell align=left,
]

\addplot[
    fill=myblueRB,
    draw=myblueborderRB,
    line width=0.4pt,
] table[x=model_short, y=tunnel_pct] {\bardata};
\addlegendentry{P0}

\addplot[
    fill=mygreenRB,
    draw=mygreenborderRB,
    line width=0.4pt,
] table[x=model_short, y=delta_baseline] {\bardata};
\addlegendentry{P1}

\addplot[
    fill=mygrayRB,
    draw=mygrayborderRB,
    line width=0.4pt,
] table[x=model_short, y=delta_strongly] {\bardata};
\addlegendentry{P2}

\addplot[
    fill=myredRB,
    draw=myredborderRB,
    line width=0.4pt,
] table[x=model_short, y=delta_force_f] {\bardata};
\addlegendentry{P3}

\end{axis}
\end{tikzpicture}
\caption{Ablation study: Latency overhead across configurations P0--P3 (see \cref{tab:config_features} for feature details) for various computer-vision models. 
The y-axis shows relative latency compared to the best-case (P0), with each bar segment representing the contribution of a specific optimization feature.
}
\label{fig:ablation}
\end{figure}

\begin{table}[t]
\centering
\caption{Configuration features enabled for each optimization level (P0--P3) used in \cref{fig:ablation}.}
\label{tab:config_features}
\footnotesize
\setlength{\tabcolsep}{2pt}
\renewcommand{\arraystretch}{1.1}
\begin{tabular}{lccc}
\toprule
 & \textbf{Layer fusion} & \textbf{Hide \ac{DM}} & \textbf{Both Parallelisms} \\\midrule
P0 & \checkmark & \checkmark & \checkmark \\ 
P1 & \texttimes & \checkmark & \checkmark \\ 
P2 & \texttimes & \texttimes & \checkmark \\ 
P3 & \texttimes & \texttimes & \texttimes \\\bottomrule
\end{tabular}
\end{table}

\subsection{LLM Inference}\label{sec:llm}
\highlightRIII{
Although primarily designed for CNN inference, the proposed architecture provides sufficient flexibility to accelerate modern LLM workloads. The NPU itself supports integer arithmetic, while our LLM deployment employs a state-of-the-art hybrid quantization approach~\cite{xiao2023smoothquant}, dynamically quantizing MatMul operations while retaining the remaining operations in floating point to preserve model quality. The integrated NPU therefore collaborates with the host CPUs (6$\times$ Cortex-A55 at \SI{1.8}{GHz} with NEON-SIMD), with the NPU accelerating the quantized operations and the CPUs executing the remaining floating-point operations.

As the reference NPU systems used for the CNN comparison do not support the evaluated hybrid-quantized LLMs, Table~\ref{tab:cpu_npu_comparison} instead compares the proposed NPU+CPU configuration against CPU-only execution. The CPU-only baseline uses the same six Cortex-A55 cores and the same \SI{12.5}{GB/s} DDR bandwidth limit as the NPU+CPU configuration. The NPU reduces time to first token (TTFT) by up to 67.5\% (TinyLlama-1.1B, sequence length 128), as the compute-bound prefill phase benefits from NPU acceleration~\cite{verhelst2025keep}. For the memory-bound decode phase, the NPU increases token throughput by up to 32.0\% (TinyLlama-1.1B, sequence length 64) by exploiting the available DDR bandwidth and lossless compression~\cite{verhelst2025keep}.

Overall, the results demonstrate that the proposed system provides meaningful performance gains for the evaluated hybrid-quantized LLM workloads. The relative gains are smaller than for the CNN workloads, for which end-to-end speedups over CPU-only inference can exceed $30\times$.
}

\begin{table}[t] 
    \centering 
    \caption{CPU and NPU+CPU LLM inference performance comparison.} 
    \label{tab:cpu_npu_comparison} 
    \scriptsize 
    \setlength{\tabcolsep}{2.2pt} 
    \renewcommand{\arraystretch}{1.15} 
    \resizebox{\columnwidth}{!}{ 
        \begin{tabular}{l c cc cc cc} 
            \toprule 
            \multirow{4}{*}{\textbf{Model}} & 
            \multirow{4}{*}{\makecell{\textbf{Seq.}\\\textbf{len.}}} & 
            \multicolumn{2}{c}{\textbf{CPU}} & 
            \multicolumn{2}{c}{\textbf{NPU+CPU}} & 
            \multirow{4}{*}{\makecell{\textbf{$\Delta$TTFT}\\\textbf{(NPU} \\ \textbf{vs} \\ \textbf{CPU)}}} & 
            \multirow{4}{*}{\makecell{\textbf{$\Delta$T/s}\\\textbf{(NPU} \\ \textbf{vs} \\ \textbf{CPU)}}} \\ 
            \cmidrule(lr){3-4} 
            \cmidrule(lr){5-6} 
            & 
            & \textbf{TTFT}$\downarrow$ & \textbf{T/s$\uparrow$} 
            & \textbf{TTFT$\downarrow$} & \textbf{T/s$\uparrow$} 
            & & \\ 
            & 
            & \textbf{(s)} & 
            & \textbf{(s)} & 
            & & \\ 
            \midrule
            \midrule 
            \multirow{2}{*}{TinyLlama 1.1B} 
            & 64 & 1.23 & 7.04 & 0.42 & 9.31 & $-65.9\%$ & $+32.0\%$ \\ 
            & 128 & 2.51 & 6.85 & 0.81 & 8.91 & $-67.5\%$ & $+30.0\%$ \\
            \midrule
            \multirow{2}{*}{Qwen3 0.6B} 
            & 64 & 0.71 & 8.46 & 0.45 & 10.0 & $-37.0\%$ & $+18.2\%$ \\
            & 128 & 1.36 & 7.6 & 0.79 & 8.76 & $-42.1\%$ & $+15.3\%$ \\
            \midrule
            \multirow{2}{*}{Llama3.2 1B} 
            & 64 & 1.24 & 6.36 & 0.5 & 8.16 & $-60.0\%$ & $+28.3\%$ \\ 
            & 128 & 2.5 & 6.25 & 0.98 & 7.95 & $-60.9\%$ & $+27.2\%$ \\ 
            \bottomrule
        \end{tabular}
    }
\end{table}

\subsection{Compile Time}

\highlightRI{
Partitioning the optimization problem as described in Section~\ref{subsec:opt} enables a custom solver for scheduling and allocation that further reduces compile time compared with the results reported in Table~\ref{tab:scalability}.

The custom solver replaces computationally expensive global optimization with a deterministic greedy heuristic followed by local search. It first constructs a feasible schedule using a topological scheduling pass and then incrementally improves it through three targeted optimization passes for transfer overlap, format-conversion hiding, and weight reuse. By using greedy best-first moves instead of exploring the global optimization space, the solver substantially reduces schedule-generation time while retaining most of the inference-performance benefits.

Across the evaluated workloads, the Custom Solver reduces compilation time by 93.8\% on average, with all evaluated compilation tasks completing in under one minute on a standard desktop machine, while increasing end-to-end inference latency by only 0.78\% on average compared with the original solver.
}

\section{Conclusion and Future Work}
\label{sec:conclusions}

This work presented a commercial-grade near-memory-compute \ac{NPU} architecture and constraint-programming-based compiler mid-end for efficient edge inference. The proposed hardware--software co-design minimizes data movement and improves compute utilization under constrained memory and bandwidth resources. Despite providing only \SI{2}{TOPS} of peak compute, the system achieves an average \SI{1.8}{\times} speedup over an embedded NPU with comparable resources and \SI{1.25}{\times} over an \SI{11}{TOPS} AI-Vision SoC. These results demonstrate that peak TOPS alone is a poor predictor of end-to-end inference performance and that tightly coupled hardware--software optimization can achieve lower latency with substantially less provisioned peak compute.

Beyond the evaluated CNN workloads, the proposed system also accelerates hybrid-quantized LLM inference, reducing time to first token by up to 67.5\% over CPU-only execution. Future work will extend the heterogeneous execution model to enable fine-grained, concurrent execution of integer and floating-point operations, further improving utilization across NPU and floating-point compute resources.

\section*{Acknowledgments}
This result is part of the IPCEI ME/CT and is funded by the European Union Next Generation EU, the German Federal Ministry for Economic Affairs and Energy, the Bavarian Ministry of Economic Affairs, Regional Development and Energy, the Free State of Saxony with the help of tax revenue based on the budget approved by the Saxon State parliament and the Free and Hanseatic City of Hamburg.


\bibliographystyle{IEEEtran}
\bibliography{refs}

\newpage
\section*{Biography Section}
\vskip -2\baselineskip plus -1fil
\begin{IEEEbiography}[{\includegraphics[width=1in,height=1.25in,clip]{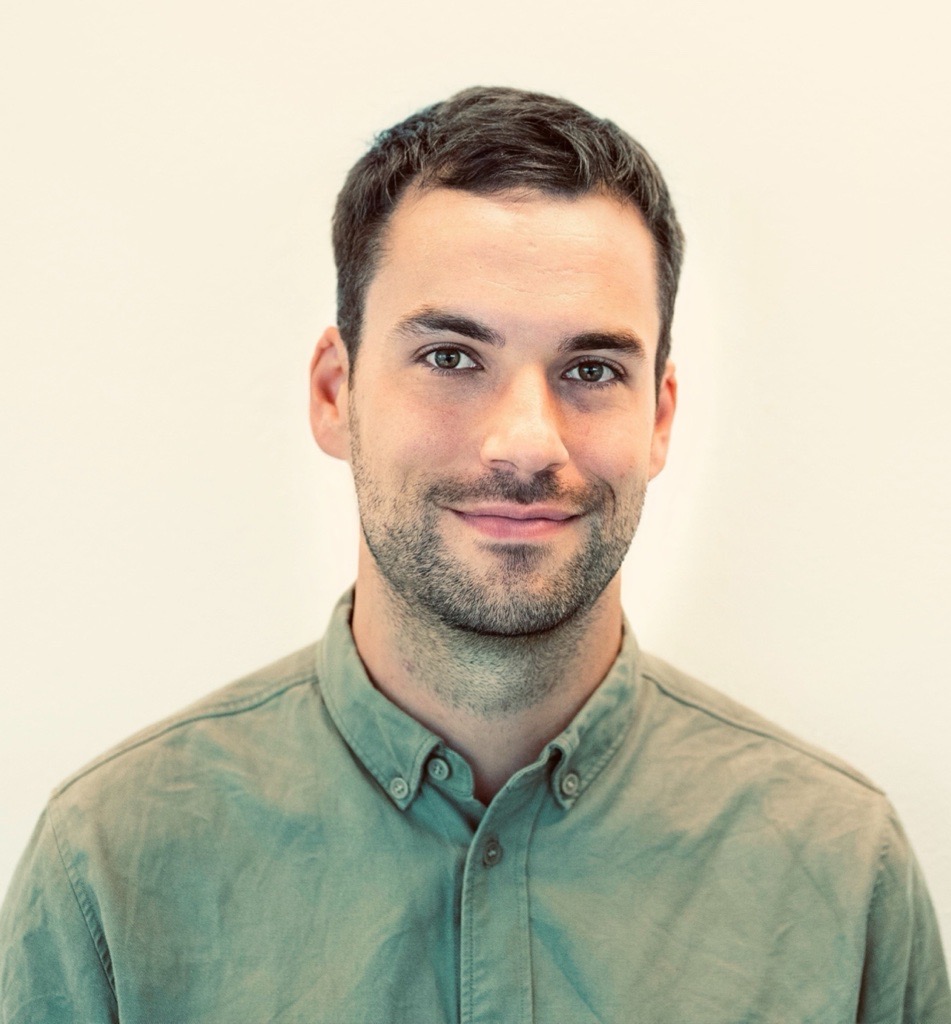}}]{Lennart Bamberg}
received the Ph.D. degree with \textit{summa cum laude} honors in Electrical and Computer Engineering from the University of Bremen, Germany. He is the Global R\&D Director for AI Systems at NXP Semiconductors, Hamburg, Germany, where he previously served as Chief NPU Architect. His research interests include domain-specific architectures for machine learning, hardware--software co-design, and resource-efficient computing systems. He has authored numerous publications, holds several patents, and has received multiple Best Paper Awards for his work. Dr. Bamberg is appointed Professor and Head of Digital Systems and Integrated Circuits at Paderborn University, Germany. He serves as Deputy Editor-in-Chief and Area Chair of \textit{IEEE Embedded Systems Letters}.
\end{IEEEbiography}
\vskip -2\baselineskip plus -1fil
\begin{IEEEbiography}[{\includegraphics[width=1in,height=1.25in,clip]{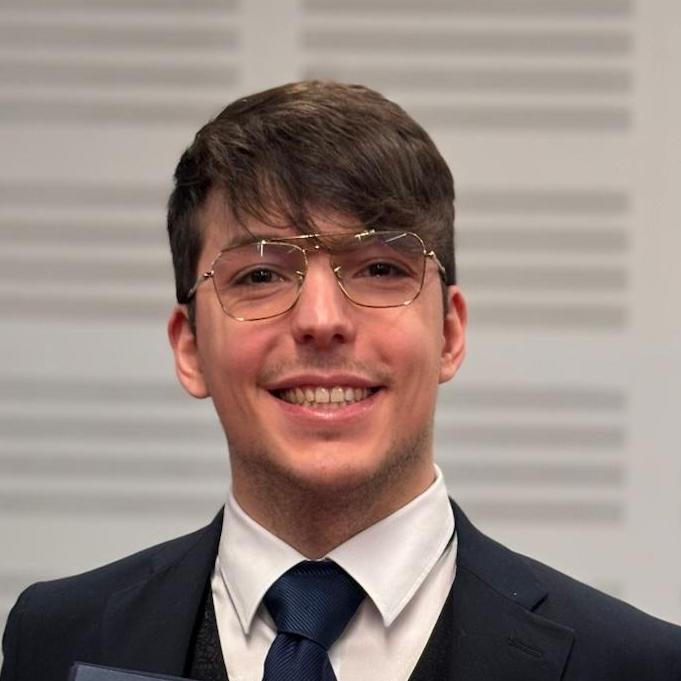}}]{Filippo Minnella} received the Ph.D. in Electrical, Electronics and Communications Engineering from Politecnico di Torino in 2024. 
He worked for STMicroelectronics Automotive group as IC Designer and for NXP Semiconductors as a Principal NPU Architect. 
He is currently working at Nvidia as Principal AI Processor Architect.
\end{IEEEbiography}%
\vskip -2\baselineskip plus -1fil
\begin{IEEEbiography}[{\includegraphics[width=1in,height=1.25in,clip]{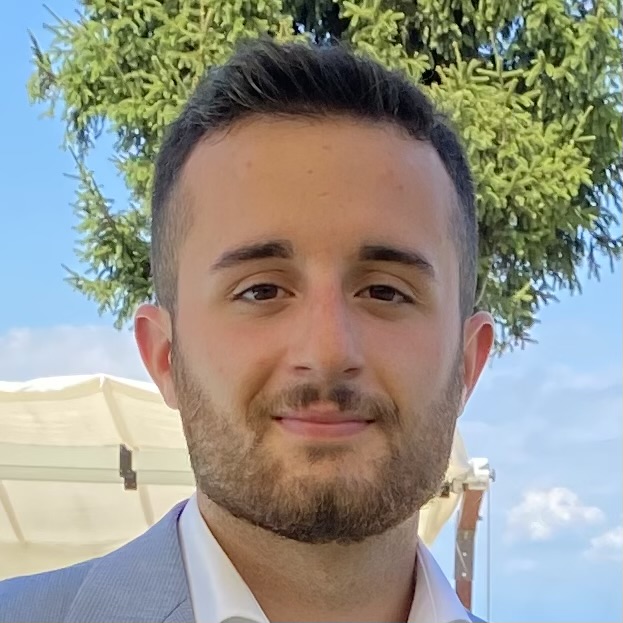}}]{Roberto Bosio} received the Master's Degree from Politecnico di Torino in 2022. He started his PhD at Politecnico di Torino in the following year. His main interests are High-Level Synthesis and dataflow architectures.
\end{IEEEbiography}%
\vskip -2\baselineskip plus -1fil
\begin{IEEEbiography}[{\includegraphics[width=1in,height=1.25in,clip]{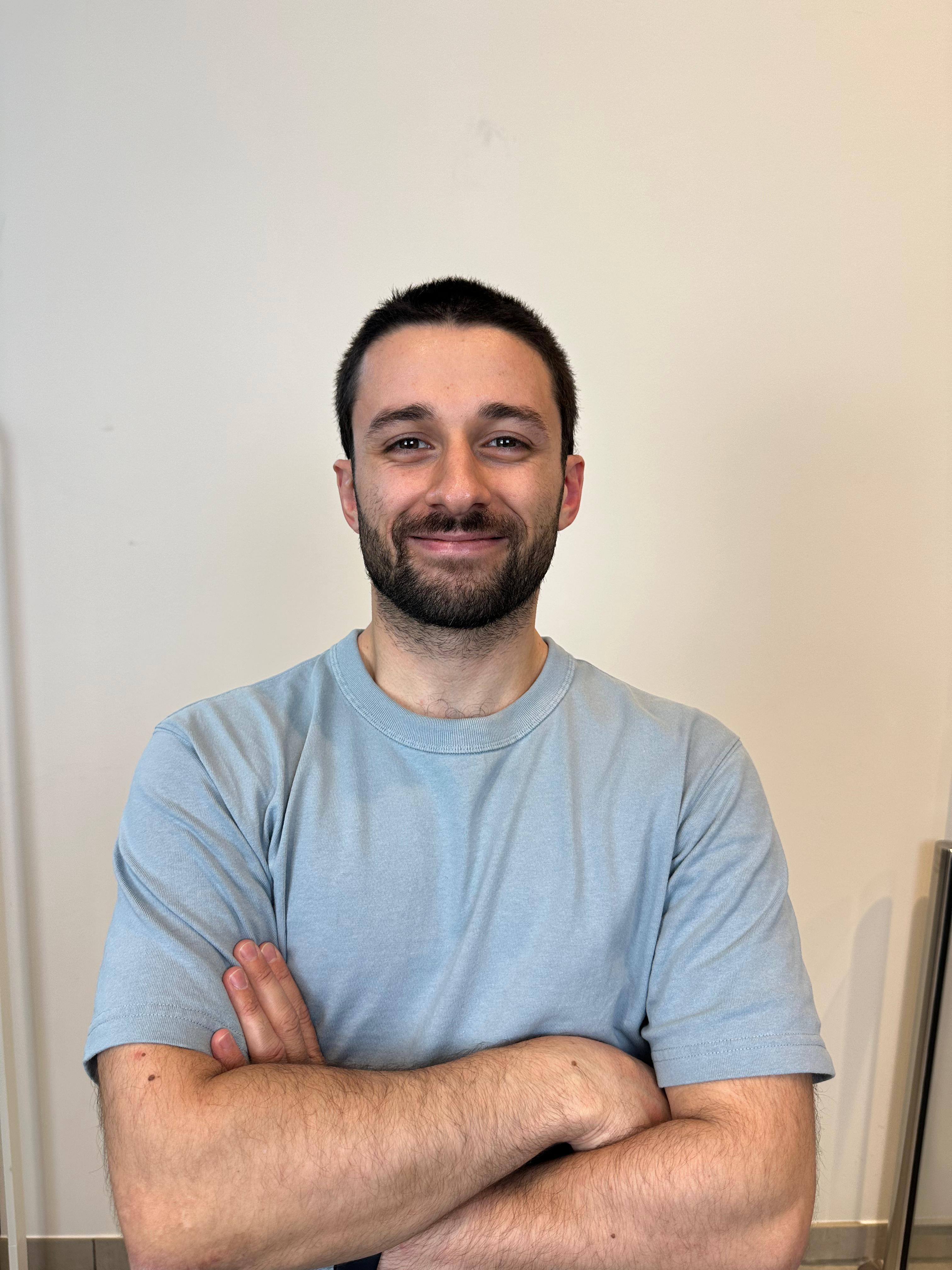}}]{Fabrizio Ottati} received his Ph.D. in Electronic Engineering from Politecnico di Torino in 2023. He is currently a Senior Computer Architect at NXP. His research interests include hardware/software codesign of machine learning accelerators and systems. 
\end{IEEEbiography}%
\vskip -2\baselineskip plus -1fil
\begin{IEEEbiography}[{\includegraphics[width=1in,height=1.25in,clip]{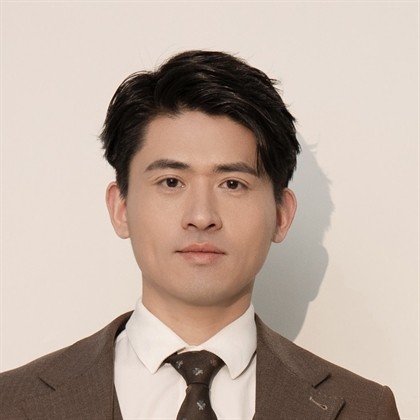}}]{Yuebin Wang} received the B.S. degree in communication engineering from Soochow University, Suzhou, China, in 2016, and the M.S. degree and Ph.D. degree in electrical engineering from Fudan University, Shanghai, China, in 2019 and 2023. Currently, he is a Principle Processor Architect with NXP Semiconductors in Eindhoven, Netherlands. His research interests include Edge-AI and signal processing.
\end{IEEEbiography}%
\vskip -2\baselineskip plus -1fil
\begin{IEEEbiography}[{\includegraphics[width=1in,height=1.25in,clip]{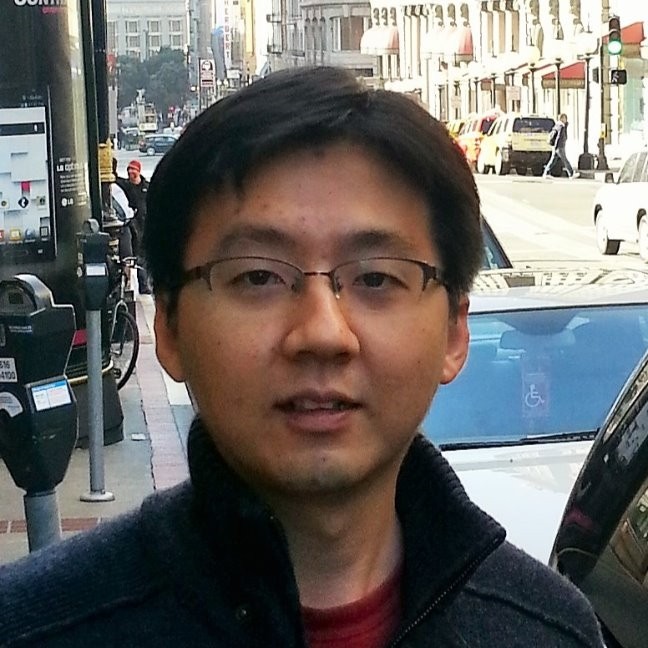}}]{Jongmin Lee} is a Principal Machine Learning Architect at NXP Semiconductors in Austin, TX, USA. Throughout his career at NXP, he has primarily focused on developing MCU-based machine learning solutions and is currently advancing NPU architecture. He earned his Ph.D. in Electrical Engineering from Arizona State University in 2017. He holds a B.S. in Electrical Communications Engineering from Konkuk University and an M.S. from the Korea Advanced Institute of Science and Technology (KAIST). Prior to his doctoral studies, he was a Research Engineer at KAIST Institute, where he contributed to several patents.
\end{IEEEbiography}%
\begin{IEEEbiography}[{\includegraphics[width=1in,height=1.25in,clip]{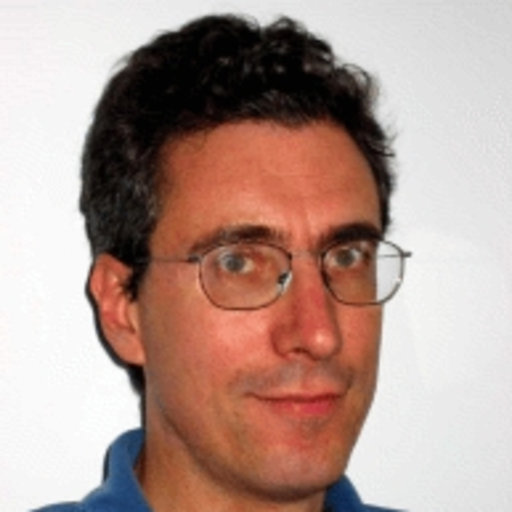}}]{Luciano Lavagno} (SM’89) received the Ph.D. degree in electrical engineering and computer science from U.C. Berkeley in 1992. He was an Architect with the POLIS HW/SW co-design tool. From 2003 to 2014, he was an Architect with the Cadence C to Silicon high-level synthesis tool. Since 1993, he has been a Professor with the Politecnico di Torino, Italy. He co-authored four books and over 200 scientific papers. His research interests include synthesis of asynchronous circuits, HW/SW codesign and high-level synthesis. 
\end{IEEEbiography}%
\newpage
\begin{IEEEbiography}[{\includegraphics[width=1in,height=1.25in,clip]{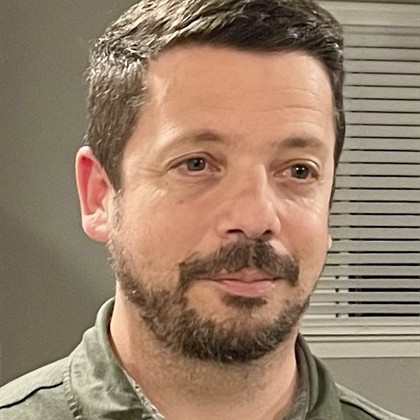}}]{Adam Fuks} 
 is a Fellow in Processor Architecture at NXP Semiconductors, based in San Jose, CA, USA. He has over two decades of experience in microcontroller and processor design, specializing in low-power architectures, security, and scalable compute subsystems for edge and embedded systems. At NXP, he drives architectural strategy and innovation across heterogeneous compute platforms and leads initiatives for next-generation AI-enabled processors. Adam holds numerous patents in processor and system architecture and has played a key role in defining industry-leading microcontroller platforms.
\end{IEEEbiography}

\end{document}